\input ./style/arxiv-general.cfg
\documentclass[aos,MSNbibl,nameyear,dvips]{arximspdf}
\makeatletter
   \@ifpackageloaded{graphicx}{}{\usepackage{graphicx}}
\makeatother
\usepackage{algpseudocode}
\usepackage{algorithm}

\doi{10.1214/15-AOS1329}
\volume{43}
\issue{5}
\pubyear{2015}
\firstpage{1896}
\lastpage{1928}
\docsubty{FLA}

\makeatletter
\newcommand{\rrvert}{\vert}
\newcommand{\llvert}{\vert}
\def\cal{\mathcal}

\newcommand{\eqref}[1]{(\ref{#1})}
\newtheorem{thmm}{Theorem}
\newtheorem{lem}[thmm]{Lemma}

\newproclaim{remark}{Remark}

\newcommand\R{\mathbb{R}}
\newcommand\E{\mathbb{E}}
\renewcommand\P{\mathbb{P}}
\newcommand\Haus{\mathsf{ Haus}}
\newcommand\dist{\mathsf{ dist}}
\newcommand\rank{\mathsf{ rank}}
\newcommand\reach{\mathsf{ reach}}

\newcommand\cL{{\cal L}}

\newcommand{\Cov}{\operatorname{Cov}}
\newcommand{\Ridge}{\operatorname{Ridge}}

\def\hat{\widehat}
\def\tilde{\widetilde}

\makeatother

\begin{document}
\begin{frontmatter}

\title{Asymptotic theory for density ridges}
\runtitle{Asymptotic ridges}

\begin{aug}
\author[A]{\fnms{Yen-Chi}~\snm{Chen}\corref{}\ead[label=e1]{yenchic@andrew.cmu.edu}\thanksref{T1}},
\author[A]{\fnms{Christopher R.}~\snm{Genovese}\ead[label=e2]{genovese@cmu.edu}\thanksref{T2}}
\and
\author[A]{\fnms{Larry}~\snm{Wasserman}\ead[label=e3]{larry@cmu.edu}\thanksref{T3}}
\runauthor{Y.-C. Chen, C.~R. Genovese and L. Wasserman}
\affiliation{Carnegie Mellon University}
\thankstext{T1}{Supported by DOE Grant  DE-FOA-0000918.}
\thankstext{T2}{Supported by DOE Grant  DE-FOA-0000918
and NSF Grant  DMS-12-08354.}
\thankstext{T3}{Supported by NSF Grant  DMS-12-08354.}

\address[A]{Department of Statistics\\
Carnegie Mellon University\\
5000 Forbes Ave.\\
Pittsburgh, Pennsylvania 15213\\
USA\\
\printead{e1}\\
\phantom{E-mail:\ }\printead*{e2}\\
\phantom{E-mail:\ }\printead*{e3}}
\end{aug}

%
\received{\smonth{6} \syear{2014}}
%
\revised{\smonth{3} \syear{2015}}

%
\begin{abstract}
The large sample theory of estimators for density modes is well understood.
In this paper we consider density ridges, which are a higher-dimensional
extension of modes.
Modes correspond to zero-dimensional, local high-density regions in
point clouds.
Density ridges correspond to $s$-dimensional, local high-density
regions in point clouds.
We establish three main results.
First we show that under appropriate regularity conditions,
the local variation of the estimated ridge
can be approximated by an empirical process.
Second, we show that the distribution of the estimated ridge converges
to a Gaussian process.
Third, we establish that the bootstrap leads to valid
confidence sets for density ridges.
\end{abstract}

%
\begin{keyword}[class=AMS]
\kwd[Primary ]{62G20}
\kwd[; secondary ]{62G15}
\kwd{62G05}
\end{keyword}

\begin{keyword}
\kwd{Ridge}
\kwd{density estimation}
\kwd{nonparametric statistics}
\kwd{empirical process}
\kwd{bootstrap}
\end{keyword}
%
\end{frontmatter}

\section{Introduction}\label{sec1}

There is a large literature on the problem
of estimating the modes of a density.
Known results include minimax rates of convergence,
limiting distributions
and the validity of bootstrap inference
[\citeauthor{romano1988weak} (\citeyear{romano1988weak,romano1988bootstrapping})].
The purpose of the current paper is to establish
similar results for the estimation of \emph{ridges},
an extension of modes to higher dimensions.

\begin{figure}
\centering
\begin{tabular}{@{}cc@{}}

\includegraphics{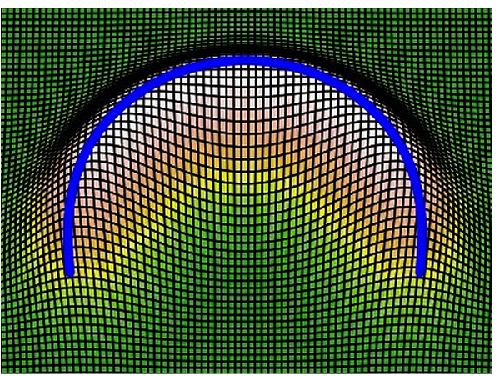}
 & \includegraphics{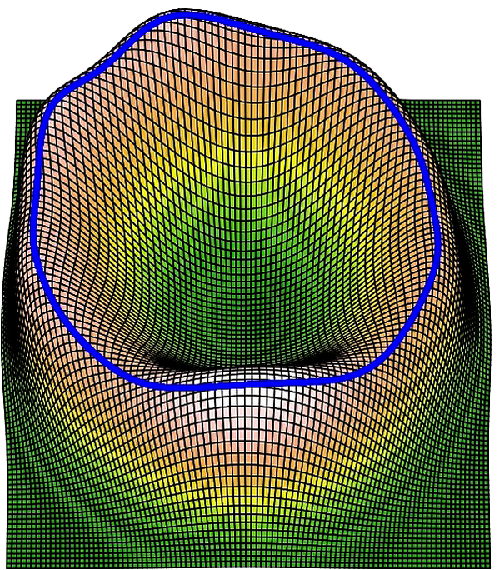}\\
\footnotesize{(a)} & \footnotesize{(b)}
\end{tabular}
\caption{Examples for the density ridges.}\vspace*{-5pt}
\label{fig:RidgeEstimation}
\end{figure}

\begin{figure}[b]\vspace*{-5pt}
\centering
\begin{tabular}{@{}cc@{}}

\includegraphics{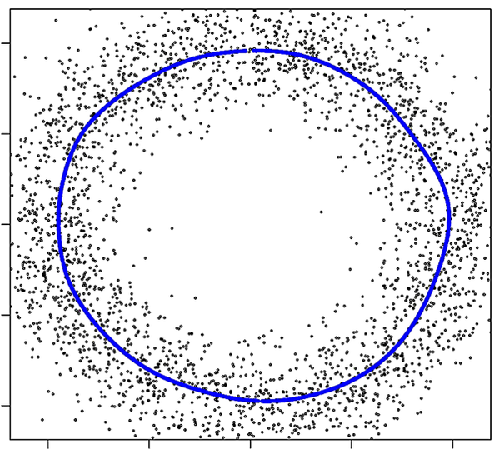}
 & \includegraphics{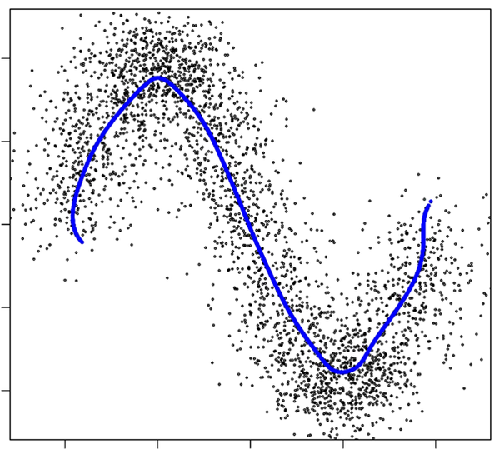}\\
\footnotesize{(a)} & \footnotesize{(b)}
\end{tabular}
\caption{Examples of estimated ridges (blue curves).}
\label{Fig::EX2}
\end{figure}

Intuitively, an $s$-ridge of a density
is an $s$-dimensional set of high-density concentration.
Modes are just $0$-ridges.
A density's ridges provide a useful summary of its structure
and are features of interest in a variety methods and applications.
Figure~\ref{fig:RidgeEstimation} shows some one-dimensional density ridges.
Figure~\ref{Fig::EX2} shows two simple datasets and estimates of the ridges.

In this paper, we consider the $s=1$ case,
and we study the large-sample behavior of
the plug-in ridge estimator based on a kernel
estimator for the underlying density.
Let $p$ be a density, and let $\hat p_h$ be a kernel estimator
with bandwidth $h$.
The mean $p_h = \mathbb{E}(\hat p_h)$ is a smoothed version of the density.
We let $R = \operatorname{ Ridge}(p)$ denote the ridge of a density $p$,\vspace*{1pt}
defined formally in Section~\ref{sec::ridges}.
We define $\hat{R}_h\equiv\Ridge(\hat{p}_n)$ as the estimated ridge and
$R_h\equiv\Ridge(p_h)$ as the smoothed ridge.

We focus on $R_h$ rather than $R$ for three reasons.
First, there is an unavoidable bias in estimating $R$ by $\hat{R}_h$.\vadjust{\goodbreak}
This bias originates intrinsically from the kernel density estimator (KDE).
In contrast, estimating $R_h$ is unbiased, which allows us to focus on
the stochastic variation of $\hat{R}_h$.
Second, as is shown in \citet{Genovese2012a},
when a topological assumption called tameness is assumed
[\citet{cohen2007stability}, \citeauthor{chazal2009proximity}
(\citeyear{chazal2009proximity,chazal2012structure})],
then $R_h$ and $R$ have the same topology
for small $h$.
In addition, for fixed $h$, the convergence rate for estimating
$R_h$ is fast.
The third reason is that when the kernel is smooth, $R_h$ is always
well defined,
while
$R$ may be nonsmooth or may not even exist.
The main results of the paper focus on characterizing the uncertainty
in the ridge estimator.
Here is a summary of the main results:

\textit{Result \textup{1:} Local uncertainty of ridges
\textup{(}Theorem~\ref{LU2}\textup{)}}.
Let $\pi_A(x)$ be the projection of $x$ onto a set $A$.
We define the local uncertainty as the vector
$\mathbf{d}(x,\hat{R}_h) = \pi_{\hat{R}_h}(x)-x$
for $x\in R_h$.
Note that this vector is unique
when $\hat{R}_h$ and $R_h$
are close and $\hat{R}_h$ is sufficiently smooth.\vadjust{\goodbreak}
We show that
the local uncertainty
can be approximated by an empirical process $\mathbb{G}_n$,
%
\begin{equation}
\sup_{x\in R_h}\bigl\|\sqrt{nh^{d+2}}
\mathbf{d}(x,\hat {R}_h)-\mathbb {G}_n(f_x)
\bigr\|_{\infty} = O_P \biggl(\sqrt{\frac{\log n}{nh^{d+6}}} \biggr)
\label{eq::Main1}
\end{equation}
for some class of
functions $y\mapsto f_x(y) \in\mathbb{R}^d$. We use
%
\begin{equation}
\rho^2_n(x)= \mathbb{E} \bigl(d(x,\hat{R}_h)^2
\bigr) = \mathbb{E} \bigl(\bigl\|\mathbf{d}(x,\hat{R}_h)\bigr\|^2
\bigr),\qquad x\in R_h,
\end{equation}
to measure the local uncertainty.
Note this quantity is essentially a local mean squared error.

\textit{Result \textup{2:} Limiting distribution
\textup{(}Theorem~\ref{thmm::GP}\textup{)}}.
Let $\mathbb{B}$ be a Gaussian process defined on a function space
$\mathcal{F}_h$
defined later.
If $n h^{d+8}/\log n \to\infty$, then
we have the following Berry--Esseen result:
%
\begin{eqnarray}
&&\sup_t \Bigl\llvert \P\bigl(\sqrt{nh^{d+2}}\Haus(
\hat{R}_h,R_h) \leq t\bigr) - \P \Bigl(\sup
_{f\in\mathcal{F}_h}\bigl\|\mathbb{B}(f)\bigr\| \leq t \Bigr) \Bigr\rrvert
\nonumber
\\[-8pt]
\\[-8pt]
\nonumber
&&\qquad= O \biggl(
\frac{\sqrt{\log n}}{(nh^{d+2})^{1/8}} \biggr),
\end{eqnarray}
where $\Haus(A,B)$ is the Hausdorff distance between two sets $A,B$.

\textit{Result \textup{3:} Bootstrap validity \textup{(}Theorems~\ref{BT},
\ref{thmm::GPB}, \ref{thmm::GP5}\textup{)}}.
Given $h \equiv h_n$ satisfying $n h^{d+8}/\log n \to\infty$,
the bootstrap gives valid estimates of uncertainty in three senses.
First, the local uncertainty measures $\rho^2_n(x)$ can be estimated
by the bootstrap.
Second, the distribution of Hausdorff distance
$\Haus(\hat{R}_h,R_h)$ can be estimated by the bootstrap, in the
sense that
\[
\sup_t \bigl|\hat F(t) - F(t)\bigr| = O_P \biggl(
\frac{\sqrt{\log
n}}{(nh^{d+2})^{1/8}} \biggr),
\]
where
\begin{eqnarray*}
\hat F(t) &=& \P \bigl(\sqrt{nh^{d+2}}\Haus\bigl(\hat{R}^*_h,
\hat{R}_h\bigr) \le t | X_1,\ldots, X_n
\bigr),
\\
F(t) &=& \P \bigl(\sqrt{nh^{d+2}}\Haus(\hat{R}_h,R_h)
\le t \bigr),
\end{eqnarray*}
where $\hat R_n^*$
is constructed from
$X_1^*,\ldots, X_n^*$ drawn i.i.d. from the empirical distribution $\P_n$.
And third, a bootstrap confidence set is consistent, as
%
\begin{equation}
\P\bigl(R_h \in\hat R_h \oplus\varepsilon_\alpha^*
\bigr) = 1 - \alpha+ O \biggl(\frac{\sqrt{\log n}}{(nh^{d+2})^{1/8}} \biggr),
\end{equation}
where
$\varepsilon_\alpha^*$ is an appropriate bootstrap quantile
and
\[
A \oplus\varepsilon= \bigl\{x\in\mathbb{R}^d\dvtx d(x,A)\leq
\varepsilon \bigr\}
\]
denotes the union of $\varepsilon$-balls centered on points in $A$.

\textit{Related work}. Much early work on ridge estimation
focused on image analysis [\citet{Eberly1996,Damon1999}].
The concept
of ridges in point clouds
was introduced by \citet{Hall1992,Hall2001,Wegman2002,Cheng2004}.
An algorithm for finding density ridges was given by
\citet{Ozertem2011}.
Recently,
\citet{Genovese2012a} provided some fundamental results
on density ridge estimation,
including the convergence rate and some stability properties
of plug-in ridge estimators.
A similar but distinct concept called \textit{ridgelines} was
introduced by
\citet{Ray2005} and \citet{Li2007} for Gaussian mixture models.
Metric graph reconstruction~[\citet{Aanjaneya2012,Lecci2013}],
a~method based on computational geometry, is another method
for modeling ridge structure in a point cloud.
This method tends to work best when the data are highly concentrated
along filamentary
structures and there is little noise.
An alternative approach based on minimizing sums of squares subject to
a penalty function is proposed by
\citet{Dejan}; the statistical properties of this approach are not known.
The contour tree [level set tree; \citet
{klemela2004visualization,zaliapin2012tokunaga}]
is a similar method,
but it uses high-density level sets to summarize the distribution
rather than ridges.

Ridge estimation is a branch of geometric statistics.
Limiting distributions in geometric statistics
often involve
the Hausdorff distance
[\citet{Molchanov2005}].
Examples of using the Hausdorff
distance appear in estimating density
level sets~[\citet{Tsybakov1997,Walther1997,Cuevas2006,Rinaldo2010a}],
curves [\citet{Lee1999,Cheng2005}], filaments
[\citeauthor{Genovese2010} (\citeyear{Genovese2010,Genovese2012a})] and manifolds
[\citeauthor{Genovese2012b} (\citeyear{Genovese2012b,Genovese2012c})].

In a recent paper, \citet{Qiao2014} give another asymptotic analysis for
density ridges (called filaments in that paper).
Their approach is quite different;
they prove an extreme value distribution
as the limiting result for estimating gradient ascent.
Also, they focus on the case $d=2$.

\textit{Outline}.
We begin with a formal definition of density
ridges and ridge estimators in Section~\ref{sec::background}. Then
we define the local uncertainty and confidence sets for the
density ridges. Section~\ref{sec::result} contains our main results.
We first show in Section~\ref{sec::NS} that for each point on the
ridge, we can define a $d-1$ dimensional subspace normal to the
ridge. We show that the local uncertainty
can be coupled with an empirical process (Section~\ref{sec::LU}). This
leads to the Gaussian approximation for
the Hausdorff distance (Section~\ref{sec::GP}). Finally, we prove
the consistency of the bootstrap for constructing the confidence sets
(Section~\ref{sec::GPB}). Some simple simulation results are given in
Section~\ref{sec::example}.

Throughout the paper,
we use $d$ for the dimension of ambient space and $s$ for the dimension
of ridge.
Also, we use $\hat{p}_h$ for the
kernel density estimator
and $p_h$ for the mean of $\hat p_h$.\vadjust{\goodbreak}


\section{Background}
\label{sec::background}

\subsection{Density ridges}
\label{sec::ridges}

Let $X_1,\ldots, X_n$ be random sample from a distribution $P$ with compact
support in $\R^d$ with density $p$. Let $g(x) = \nabla p(x)$ and
$H(x)$ denote the gradient and Hessian, respectively, of $p(x)$. We
begin by defining the \emph{ridges} of~$p$, as
in~\citet{Genovese2012a,Ozertem2011,Eberly1996}.

While there are many
possible definitions of ridges, this definition
has many useful properties, including
stability to perturbations in the underlying density,
estimability at a good rate of convergence and
fast reconstruction algorithms, as described in \citet{Genovese2012a}.

Let $v_1(x),v_2(x),\ldots,v_d(x)$
denote the eigenvectors
of the Hessian matrix $H(x)$
corresponding to eigenvalues
$\lambda_1(x) \ge\lambda_2(x)\ge\cdots\ge\lambda_d(x)$.
Let $V_s(x) = [v_{s+1}(x)\cdots v_d(x)]$,
a $d\times(d-s)$ matrix.
We define the order-$s$ \emph{projected gradient} $G_s(x)$ by
%
\begin{equation}
G_s(x) = V_s(x)V_s(x)^T g(x).
\end{equation}
The \emph{$s$-ridge} is the collection of points
%
\begin{equation}
R \equiv R_s = \bigl\{x\dvtx G_s(x)=0,
\lambda_{s+1}(x)<0 \bigr\}.
\end{equation}
It follows that the $0$-ridge, $R_0$, is the set of local modes.
Under weak conditions, an $s$-ridge is an $s$-dimensional manifold
by the implicit function theorem.

From this point forward, we focus on the case $s=1$,
henceforth writing $G(x)=G_1(x)$ and $V(x)=V_1(x)$.
Thus
%
\begin{equation}
V(x) = \bigl[v_2(x) \cdots v_d(x)\bigr]\quad \mbox{and}\quad G(x) =
V(x)V(x)^T g(x).
\end{equation}
The $1$-ridge (or simply ridge) is thus
%
\begin{equation}
\label{eq::ridge-def} R \equiv\Ridge(p) = \bigl\{x\dvtx G(x) = 0, \lambda_2(x)
< 0 \bigr\}.
\end{equation}
We use ``ridge'' as an operator that
maps a density function to the ridge set.
Because the columns of $V(x)$ are orthonormal,
%
\begin{equation}
G(x) = 0 \quad\Longleftrightarrow \quad V(x)^T g(x)=0. \label{eq::PG}
\end{equation}
Intuitively, at points on the ridge,
the density curves sharply downward in all but the
direction of first eigenvector (corresponding
to the eigenvector of the largest eigenvalue) and along
the first eigenvector, the density
curves more gently.
By the implicit function theorem, the ridges are $1$-dimensional manifolds
if (i.e., a~collection of curves)
\[
\rank \bigl(\nabla \bigl(V(x)^T g(x) \bigr) \bigr)=d-1\qquad \forall x\in
R.
\]
Claim~4 of Lemma~\ref{lem::NS} gives a sufficient condition for the ridges
to be $1$-dimensional manifolds.

\subsection{Estimated ridges and smoothed density ridges}

Given data
$X_1,\ldots,X_n$ drawn i.i.d. from density $p$, we estimate $\Ridge
(p)$ by
%
\begin{equation}
\hat{R}_h = \Ridge(\hat{p}_h),
\end{equation}
where the kernel density estimator (KDE) $\hat p_h$ is defined by
%
\begin{equation}
\hat{p}_h(x) = \frac{1}{nh^d} \sum_{i=1}^n
K \biggl(\frac{\|x-X_i\|
}{h} \biggr).
\end{equation}
Here, the kernel $K$ is a smooth, symmetric density function such as a Gaussian
and the bandwidth $h\equiv h_n > 0$.
Figure~\ref{fig:RidgeEstimation} shows an example.

\begin{figure}

\includegraphics{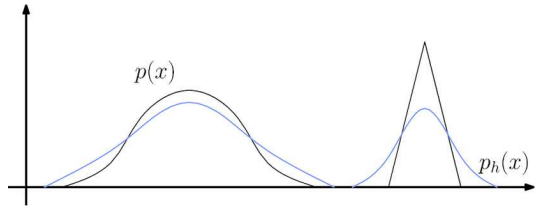}

\caption{An example of a density $p(x)$ (black curves) versus the
smoothed density $p_h(x)$ (blue curves). Note that even if $p(x)$
is nondifferentiable, its smoothed version is very smooth provided
the kernel function is smooth.}
\label{Fig::SM}
\end{figure}

\begin{figure}
\centering
\begin{tabular}{@{}cc@{}}

\includegraphics{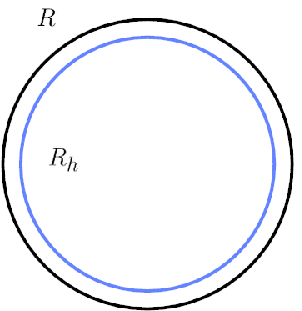}
 & \includegraphics{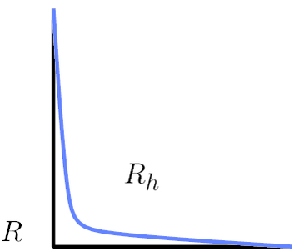}\\
\footnotesize{(a)} & \footnotesize{(b)}
\end{tabular}
\caption{Examples for ridge $R$ (black) and its smoothed
version $R_h$ (blue). Note that in \textup{(b)}, the original ridge
is nonsmooth due to the sharp angle, but the smoothed ridge
is smooth if the kernel function is
smooth enough.}
\label{Fig::RS}
\end{figure}

Let $p_h$ denote the expected value of the estimated density
$p_h(x) = \mathbb{E}(\hat{p}_h(x))$.
Thus $p_h = p\star K_h$,
where $\star$ denotes convolution.
Hence $p_h$ is a smoothed version of $p$.
Figure~\ref{Fig::SM} compares a density $p$ and its smoothed version $p_h$.
We define the \emph{smoothed ridge} set to be the ridge
set of $p_h$:
%
\begin{equation}
R_h = \Ridge(p_h) = \bigl\{x\dvtx V_h(x)^Tg_h(x)=0
\bigr\}, \label{eq::SR}
\end{equation}
where $V_h(x)=[v_2(x) \cdots v_d(x)]$, $v_k(x)$ is the eigenvector
of Hessian matrix of $p_h(x)$ corresponding to the $k$th eigenvalue
and $g_h(x) = \nabla p_h(x)$.
Figure~\ref{Fig::RS} compares the
smoothed ridge $R_h$ and the original ridge $R$. Our main focus here
is on estimating the smoothed ridge $R_h$.

\subsection{Distance measures and functional norms}

Define the projection from one point $x$ onto a set $A$ by
%
\begin{equation}
\pi_A(x) = \mathop{\operatorname{ argmin}}_{y\in A} \|x-y\|.
\end{equation}
We define the projection vector from $x$ onto a set $A$ as
%
\begin{equation}
\mathbf{d}(x,A) = \pi_A(x)-x. \label{eq::PR1}
\end{equation}
The projection vector may not be unique.
A condition related to the uniqueness of the projection is called the
\emph{reach}
and will be formally introduced in Section~\ref{sec::AS}.
The projection distance from $x$ onto $A$ is
%
\begin{equation}
d(x,A) = \bigl\|\mathbf{d}(x,A)\bigr\|. \label{eq::PR2}
\end{equation}

The Hausdorff distance between two subsets of $\R^d$
is defined by
%
\begin{equation}
\Haus(A,B) = \inf\{\varepsilon> 0\dvtx A \subset B \oplus \varepsilon \mbox{ and
}B \subset A \oplus\varepsilon\}, \label{eq::Haus}
\end{equation}
where $A \oplus\varepsilon= \bigcup_{x\in A} B(x,\varepsilon)$ and
$B(x,\varepsilon)=\{y\dvtx \|x-y\| \le\varepsilon\}$.
We also define
the
quasi-Hausdorff distance
$\dist_{\Pi}(A,B)$ as
%
\begin{equation}
\dist_{\Pi}(A,B) = \sup_{x\in B} d(x,A), \label{eq::aHaus}
\end{equation}
so that
%
\begin{equation}
B\subseteq A\oplus\dist_{\Pi}(A,B). \label{eq::CIt2}
\end{equation}
Note that
%
\begin{equation}
\Haus(A,B) = \max\bigl\{\dist_{\Pi}(A,B),\dist_{\Pi}(B,A)\bigr
\}. \label{eq::PR3}
\end{equation}

Now we introduce some norms and semi-norms characterizing the
smoothness of the density $p$. A vector $\alpha=
(\alpha_1,\ldots,\alpha_d)$ of nonnegative integers is called a
multi-index with $|\alpha| = \alpha_1 + \alpha_2 + \cdots+ \alpha_d$,
and the corresponding derivative operator is
%
\begin{equation}
D^\alpha= \frac{\partial^{\alpha_1}}{\partial x_1^{\alpha_1}} \cdots \frac{\partial^{\alpha_d}}{\partial x_d^{\alpha_d}}, \label{eq::d1}
\end{equation}
where $D^\alpha f$ is often written as $f^{(\alpha)}$.
For $j = 0,\ldots, 4$, define
%
\begin{equation}
\|p\|_{\infty}^{(j)} = \max_{\alpha\dvtx |\alpha| = j} \sup
_{x\in\R^d} \bigl|p^{(\alpha)}(x)\bigr|.
\end{equation}
When $j = 0$, we have the infinity norm of $p$; for $j > 0$, these are
semi-norms.
We also define
%
\begin{equation}
\|p\|^*_{\infty, k} = \max_{j=0,\ldots,k} \|p\|^{(j)}_{\infty}.
\end{equation}
It is easy to verify that this is a norm.

\subsection{Local uncertainty measures for the density ridges}

We define the \emph{local uncertainty}
by
%
\begin{eqnarray}
\label{eq::local-uncertainty} \rho^2_n(x) = \cases{ \E \bigl(
d^2(x, \hat{R}_h) \bigr), & \quad $\mbox{if }x \in
R_h,$ \vspace*{2pt}\cr
0, &\quad $\mbox{otherwise.}$}
\end{eqnarray}

We estimate the local uncertainty measure by the
bootstrap. Let $\mathbb{X}_n = \{X_1,\ldots, X_n\}$ be the given
observations. We define $\hat{R}^*_h$ as the estimated ridge based on
the bootstrap [\citet{Efron1979}] sample of $\mathbb{X}_n$. More
precisely, let $X^*_1,\ldots,X^*_n$ be a bootstrap sample from the
empirical distribution $\P_n$. Let $\hat{p}^*_h(x)$ be the KDE
based on the bootstrap sample. The bootstrap ridge is defined as
%
\begin{equation}
\hat{R}^*_h = \Ridge\bigl(\hat{p}^*_h(x)\bigr).
\end{equation}
We define
%
\begin{eqnarray}
\label{eq::e-local-uncertainty} \hat{\rho}^2_n(x) = \cases{ \E \bigl(
d^2\bigl(x, \hat{R}^*_h\bigr)|\mathbb{X}_n
\bigr), &\quad $\mbox{if $x \in\hat{R}_h$},$ \vspace*{2pt}
\cr
0 ,& \quad $\mbox{otherwise},$}
\end{eqnarray}
as the estimated local uncertainty.
Algorithm~\ref{Alg::LU} gives a pseudo-code for estimating $\rho
^2_n(x)$ by the bootstrap.

\begin{algorithm}[t]
\caption{Local uncertainty estimator}
\label{Alg::LU}
\begin{algorithmic}
\State\textbf{Input:} Data $\{ X_1,\ldots,X_n\}$.
\State1. Estimate the ridges from $\{ X_1,\ldots,X_n\}$; denote the
estimate by $\hat{R}_h$.
\State2. Generate $B$ bootstrap samples: $X^{*(b)}_1,\ldots
,X^{*(b)}_n$ for $b = 1,\ldots,B$.
\State3. For each bootstrap sample, estimate the ridges, yielding
$\hat{R}^{*(b)}_n$ for $b = 1,\ldots,B$.
\State4. For each $x\in\hat{R}_h$, calculate $\rho^{2}_{(b)}(x) =
d^2(x,\hat{R}^{*(b)}_h)$, $b=1,\ldots,B$.
\State5. Define $\hat{\rho}^2_n(x) = \operatorname{mean}\{\rho
^2_{(1)}(x),\ldots,\rho^2_{(B)}(x)\}$.

\State\textbf{Output:} $\hat{\rho}^2_n(x)$.
\end{algorithmic}
\end{algorithm}

\subsection{Confidence sets} \label{sec::CI}

For making inferences about ridges,
we focus on constructing a confidence set for $R_h$,
ignoring the bias $\Haus(R,R_h)$.
For suitable $h$,
$R_h$ has essentially the same shape as $R$
and thus serves as a useful target.

We call $\hat C_n \equiv\hat{C}(X_1,\ldots, X_n)$ a valid $(1-\alpha)$
confidence set if
%
\begin{equation}
\liminf_{n\to\infty}\P(R_h\subseteq\hat{C}_n)
\geq1-\alpha.
\end{equation}
Let $t_{\alpha}$ be the value such that
\[
\P(R_h \subseteq\hat{R}_h \oplus t_{\alpha})
\geq1-\alpha.
\]
Thus $t_{\alpha} = F^{-1}(1-\alpha)$ where
%
\begin{equation}
F(t) = \P\bigl(\dist_{\Pi}(\hat{R}_h,R_h)
\leq t\bigr).
\end{equation}
Although $t_\alpha$ is unknown,
we can estimate it by the bootstrap.
We define
$\hat t_{\alpha} = \hat F^{-1}(1-\alpha)$
where
%
\begin{equation}
\hat F(t) = \P \bigl(\dist_{\Pi}\bigl(\hat{R}_h^*,\hat
R_h\bigr) \leq t  | \mathbb{X}_n \bigr)
\end{equation}
and where
$\hat R_h^*$ is constructed from
an i.i.d. sample $X_1^*,\ldots, X_n^*$
from the empirical distribution $\hat{\P}_n$.
Algorithm~\ref{Alg::CI} provides a pseudo-code for constructing the
confidence sets and
Theorem~\ref{thmm::GP5} shows its consistency.

\begin{algorithm}[t]
\caption{Confidence sets}
\label{Alg::CI}
\begin{algorithmic}
\State\textbf{Input:} Data $\{ X_1,\ldots, X_n\}$, significance level
$\alpha$.
\State1. Estimate the ridge from $\{ X_1,\ldots,X_n\}$; denote this
by $\hat{R}_h$.
\State2. Generates bootstrap samples $\{ X^{*(b)}_1,\ldots
,X^{*(b)}_n\}$ for $b = 1,\ldots,B$.
\State3. For each bootstrap sample, estimate the ridge, call this
$\hat{R}^{*(b)}_h$.
\State4. For $i=1,\ldots, B$, calculate $t_i = \{\dist_{\Pi}(\hat
{R}^*_i,\hat{R}_h)\}$.
\State5. Let $\hat{t}_{\alpha}$ be the $\alpha$-upper quantile of
$t_1,\ldots, t_B$.
\State\textbf{Output:} $\hat{R}_h \oplus\hat{t}_{\alpha}$.
\end{algorithmic}
\end{algorithm}

\section{Main results}
\label{sec::result}

For a vector $v\in\mathbb{R}^d$, $\|v\|$ is the usual $\cL^2$
norm for
the vector, and $\|v\|_{\infty}$ is the supremum norm for $v$;
that is,
\[
\|v\|_{\infty} = \max\bigl\{\|v_1\|,\ldots,\|v_d\|\bigr\}.
\]
For a matrix $M$, let
$ \|M\|_{\max} = \max_{i,j}\|M_{ij}\|$.
When $M$ is symmetric, we define
$\|M\|_2 = \max_{v}\frac{\|M v\|}{\|v\|}$.

We define $\mathbf{C}^r$ to be the collection of $r$-times
continuously differentiable functions.
For a vector value function $f = (f_1,\ldots,f_k)\dvtx\mathbb
{R}^d\mapsto
\mathbb{R}^k$,
we define the gradient $\nabla f(x)$ as a $d\times k$ matrix given by
%
\begin{equation}
\nabla f(x) = \bigl(\nabla f_1(x),\ldots, \nabla f_k(x)
\bigr).
\end{equation}

\subsection{Assumptions}
\label{sec::AS}

We begin by defining the tangent vector $e(x)$ to $R_h$ at each $x\in R_h$.
Let
%
\begin{equation}
M(x) = \nabla \bigl(V_h(x)^T g_h(x)
\bigr),
\end{equation}
which is a $d\times(d-1)$ matrix.
We define $e(x)$ to be the eigenvector corresponding to
the largest eigenvalue of
$\mathbf{I}_d- M(x) (M(x)^TM(x) )^{-1}M(x)^T$.
As long as $M(x)$ has rank $d-1$, $e(x)$ is unique.

\begin{lem}
\label{lem::NS0}
Assume the matrix $M(x)$ has rank $d-1$.
Then $e(x)$, the first eigenvector of
\[
\mathbf{I}_d- M(x) \bigl(M(x)^TM(x)
\bigr)^{-1}M(x)^T,
\]
is tangent to $R_h$ at $x\in R_h$.
The column space of $M(x)$ is normal to $R_h$ at each $x\in R_h$.
\end{lem}

The proof can be found in the supplementary material
[\citet{chen2014ridgesupp}].
By Lemma~\ref{lem::NS0}, the vector $e(x)$ defined as above is always
tangent to $R_h$
whenever $x\in R_h$.
Later we will see in claim 4 of Lemma~\ref{lem::NS},
condition (P1) with smoothness on $p_h$ [guaranteed by conditions (K1)--(K2)]
implies Lemma~\ref{lem::NS0}.

With the above notation, we now formally describe our assumptions.
\begin{longlist}
\item[(K1)] The kernel $K$ is in $\mathbf{C}^{4}$ and $\|K\|
^*_{\infty,4}<\infty$.
\item[(K2)] Let
\[
\mathcal{K}_r = \biggl\{y\mapsto K^{(\alpha)} \biggl(
\frac
{x-y}{h} \biggr)\dvtx x\in\mathbb{R}^d, |\alpha|=r \biggr\},
\]
where $K^{(\alpha)}$ is defined in \eqref{eq::d1},
and let $\mathcal{K}^*_l = \bigcup_{r=0}^l \mathcal{K}_r$.
We assume that $\mathcal{K}^*_4$ is a VC-type class; that is,
there exist constants $A,v$ and a constant envelope $b_0$ such that
%
\begin{equation}
\sup_{Q} N\bigl(\mathcal{K}^*_4,
\cL^2(Q), b_0\varepsilon\bigr)\leq \biggl(
\frac
{A}{\varepsilon} \biggr)^v, \label{eq::VC}
\end{equation}
where $N(T,d_T,\varepsilon)$ is the $\varepsilon$-covering number for an
semi-metric set $T$ with metric $d_T$, and $\cL^2(Q)$ is the $L_2$ norm
with respect to the probability measure $Q$.

\item[(P1)]
There exist constants $\beta_0,\beta_1,\beta_2,\delta_0>0$ such that
%
\begin{eqnarray}
 \lambda_2(x)&\leq&-\beta_1,\nonumber
\\
\lambda_1(x)&\geq&\beta_0-\beta_1,
\\
\bigl\|g_h(x)\bigr\|\max_{|\alpha|=3}\bigl|p_h^{(\alpha)}(x)\bigr|
&\leq&\beta _0(\beta_1-\beta_2),\nonumber
\end{eqnarray}
for all $x\in R_h\oplus\delta_0$. We call $\delta_0$ the \emph{gap}.
Note that $V_h(x)$ defined in equation~\eqref{eq::SR}.

\item[(P2)] For each $x\in R_h$,
$|e(x)^T g_h(x)|^2\geq\frac{\lambda_1(x)}{\lambda_1(x)-\lambda_2(x)}$
where $e(x)$ is the direction of $R_h$ at point $x\in R_h$ defined in
Lemma~\ref{lem::NS0}.

\item[(P3)] Conditions (P1), (P2) hold
for all small $h$.
\end{longlist}

Now we discuss the conditions. (K1) is needed since the definition of
density ridge requires twice differentiability. We need additional
smoothness for making sure the estimated ridges are smooth.
(K2) regularizes the complexity of kernel functions and its partial derivatives.
This is
to ensure the fourth derivatives of the KDE will converge;
we need
the fourth derivative since the reach of $\hat{R}_h$ depends on the
fourth derivative of $\hat{p}_h$ by claim 7 in Lemma~\ref{lem::NS}.
Note that similar conditions to (K2)
appear in \citeauthor{Einmahl2000} (\citeyear{Einmahl2000,Einmahl2005}),
\citet{Gine2002}.
The Gaussian kernel
satisfies this condition.

(P1) is the eigen condition which
also appears in \citet{Genovese2012a}. This implies that the projected
gradient near the ridge is smooth. This leads to a
well-defined local normal coordinate along ridges; see
Lemma~\ref{lem::NS}. We require a slightly stronger condition
(existence of $\beta_2$) than \citet{Genovese2012a}.

We use (P2) to make sure the density ridge is also a generalized local
mode in the normal space;
see Lemma~\ref{LUlem3}.
Note that whenever $\lambda_1(x)<0$ for some $x\in R_h$,
$x$ must be a local mode in the normal space of $R_h$ at $x$ since
all eigenvalues are negative.
(P3) is required if we allow
$h\rightarrow0$; otherwise we do not need to assume it.
Note that if we
say a density $p$ satisfies (P1) or (P2), we mean that the
condition holds for $p_h$.

Finally, we consider the following assumption that
will not be assumed in our main result
but is useful and frequently assumed in
working lemmas.
\begin{longlist}
\item[(A1)] The density $p\in\mathbf{C}^4$
and has uniformly bounded derivatives to the fourth order.
\end{longlist}
This condition will not be assumed in our main results since conditions
(K1)--(K2) imply (A1) for $p_h$.

\subsection{The normal space for density ridges}
\label{sec::NS}

In this section, we show that under suitable conditions, for each point
$x$ on the density ridge we can construct a matrix $N(x)$ whose
columns span the normal space of the density ridge at $x$.

Let $L$ be a $d\times q$ matrix with orthonormal columns.
For such an $L$, we define the \emph{subspace derivative} by $\nabla
_{L} = L^T\nabla$,
which in turn gives the \emph{subspace gradient}
\[
g(x;L) = \nabla_{L} p(x)
\]
and the \emph{subspace Hessian}
\[
H(x;L) = \nabla_L \nabla_L p(x).
\]
Thus
$g(x;L)$ and $H(x;L)$ are the gradient and Hessian generated by the
partial derivatives along columns of $L$; this is the partial
derivative in the subspace spanned by columns of $L$. If $L$ is a
unit vector, then $\nabla_L$ is the \emph{directional derivative}
along $L$.

Now we construct a local normal coordinate for the
ridge.
Note in this subsection, all notation with subscript $q$ (e.g., $g_q,
H_q, V_q$)
denote the quantities defined for the smooth density $q$.
For any smooth density $q$,
let $g_q(x),H_q(x)$ denote the
gradient and Hessian of $q$.
For simplicity, we denote the eigenvectors and eigenvalues
of $H_q(x)$ using the same notation as before.
Let $v_1(x),\ldots, v_d(x)$ be the
eigenvectors of $H_q(x)$ corresponding to eigenvalues
$\lambda_1(x)\geq\cdots\geq\lambda_d(x)$.
As before,
the ridge set
$R_q=\operatorname{Ridge}(q)$ is defined as the collection of $x$ such that
$V_q(x)^Tg_q(x)=0$
with $\lambda_2(x)<0$.
By Lemma~\ref{lem::NS0}, the gradient of $V_q(x)^Tg_q(x)$
forms a matrix whose columns space spans the normal space to $R_q$
at each $x\in R_q$.

Define $M_q(x) = \nabla (V_q(x)^Tg_q(x) ) =
[m_2(x) \cdots m_d(x)]$ which is a $d\times(d-1)$ matrix.
\citet{Eberly1996} (page 65)
shows that
%
\begin{equation}
m_k(x) = \biggl(\lambda_k(x)
\mathbf{I}_d+ \frac{v_1(x)^T g_q(x)}{\lambda
_k(x)-\lambda_1(x)}\nabla_{v_1(x)}
H_q(x) \biggr) v_k(x),
\label{eq::NM}
\end{equation}
where $\mathbf{I}_d$ is the $d\times d$ identity matrix. The columns
of $M_q(x)$ span the normal space to $R_q$ at $x$. However, the
columns of $M_q(x)$ are not orthonormal. Thus we perform an
orthonormalization\vspace*{1pt} to $M_q(x)$ to construct $N_q(x)$ by the
following steps:
We have that
$M_q(x) = \nabla V_q(x)^Tg_q(x)$.
There exists a lower triangular matrix $L_q(x)$ such that
\[
L_q(x)L_q(x)^T = M_q(x)^T
M_q(x).
\]
We then define
%
\begin{equation}
\label{eq::NS3} N_q(x) = M_q(x) \bigl[L_q(x)^T
\bigr]^{-1}.
\end{equation}
Note that $M_q(x)$ might not be unique since the eigenvalues of $H_q(x)$
can have multiplicities.
When $H_q(x)$ has multiplicities, any choice of linearly independent
eigenvectors for $H_q(x)$
will work in the above construction.
As will be shown later, what we need is the smoothness of
$N_q(x)N_q(x)^T$ or $M_q(x)M_q(x)^T$,
which is unaffected by multiplicities.

The \emph{reach} [\citet{Federer1959}] for a set $A$, denoted by
$\reach(A)$,
is the largest real number $r$
such that each
$x\in\{y\dvtx d(y,A) \leq r\}$ has a unique projection onto $A$.
The reach measures the smoothness of a set.

\begin{lem}[(Properties of the normal space)]
\label{lem::NS}
Let $q$ be a density that satisfies \textup{(A1)} and \textup{(P1)},
and denote $R_q=\operatorname{
Ridge}(q)$.
Let $R^{\delta}_q = R_q\oplus\delta_0$ where $\delta_0$ is the gap
defined in \textup{(P1)}. Let $M_q(x),N_q(x)$ be constructed from
\eqref{eq::NS3}. Then:
\begin{longlist}[(1)]
\item[(1)] $N_q$ and $M_q$ have the same column space.
Also,
%
\begin{equation}
N_q(x)N_q(x)^T = M_q(x)
\bigl[M_q(x)^TM_q(x)\bigr]^{-1}M_q(x)^T.
\label{eq::NS5}
\end{equation}
That is, $N_q(x)N_q(x)^T$ is the projection matrix onto columns of $M_q(x)$.

\item[(2)] The columns of $N_q(x)$ are orthonormal to each other.

\item[(3)] For $x\in R_q$, the column space of $N_q(x)$ is normal to
the direction of $R_q$ at~$x$.

\item[(4)]
For all $x\in R_q$, $\rank(N_q(x))=\rank(M_q(x))=d-1$. Moreover,
$R_q$ is a $1$-dimensional manifold that contains no intersection and
no endpoints.
Namely, $R_q$ is a finite union of connected, closed curves.

\item[(5)] When $\Vert x-y\Vert$ is sufficiently small and $x,y \in
R^{\delta}_q$,
\[
\bigl\|N_q(x)N_q(x)^T-N_q(y)N_q(y)^T
\bigr\|_{\max} \leq A_0\bigl(\bigl\|q^{(3)}\bigr\|_{\infty
}+
\bigl\|q^{(4)}\bigr\|_{\infty}\bigr)^2 \|x-y\|,
\]
for some constant $A_0$.

\item[(6)] Assume $q'$ also satisfies \textup{(A1)} and \textup{(P1)} and $\|q-q'\|
^*_{\infty,3}$
is sufficiently small. Then
\[
\bigl\|N_q(x)N_q(x)^T-N_{q'}(x)N_{q'}(x)^T
\bigr\|_{\max} \leq A_1\bigl(\bigl\|q-q'\bigr\|
^*_{\infty,3}\bigr)
\]
for some constant $A_1$.

\item[(7)] The reach of $R_q$ satisfies
\[
\reach(R_q)\geq\min \biggl\{\frac{\delta_0}{2}, \frac{\beta
^2_2}{A_2(\|q^{(3)}\|_{\infty}+\|q^{(4)}\|_{\infty})}
\biggr\}
\]
for some constant $A_2$.
\end{longlist}
\end{lem}

The proof can be found the supplementary material [\citet{chen2014ridgesupp}].
We call $N_q(x)$ the \emph{normal matrix} since by claims 2 and 3 of
the lemma, the
columns of $N_q(x)$ span the normal space to $\operatorname{Ridge}(q)$ at $x$.
By claim 4, the ridge is a 1-dimensional manifold
and by claim 1--3 and Lemma~\ref{lem::NS0}, at each $x\in R_h$,
the column space of $N_q(x),M_q(x)$ spans
the normal space to $R_h$ at~$x$.

Claim~4 avoids cases in density ridges that are not well defined:
endpoints and intersections. The eigenvectors near
endpoints or intersections will be ill-defined.
Claim~5 proves that the projection matrix, $N_q(x)N_q(x)^T$,
changes smoothly near $R_q$.
Claim~6
shows that when two density functions are sufficiently close, the
column space of $N_q(x)$ will also be close.
Claim~7 gives the smoothness of $R_q$ in terms of the reach.

In the following sections, we work primarily on the ridge generated
from $p_h$ and $\hat{p}_h$, so for simplicity we define
%
\begin{equation}
N(x) = N_{p_h}(x), \qquad \hat{N}_n(x) = N_{\hat{p}_h}(x).
\label{eq::NS4}
\end{equation}

\subsection{Local uncertainty for ridges}
\label{sec::LU}
Let
\[
H_N(x) = H\bigl(x;N(x)\bigr),\qquad  x\in R_h,
\]
which is the subspace Hessian matrix in the normal space along $R_h$ at $x$.
Recall that $N(x)$ is not uniquely defined (due to possible
multiplicities of eigenvalues),
but any choice of $N(x)$ constructed from \eqref{eq::NS3} can be used
in the definition of $H_N$.
Lemma~\ref{lem::SH} guarantees this invariance.

Let ${\cal F}$ be the class of vector valued functions
defined by
%
\begin{equation}
\label{eq::LU2eq1} {\cal F} = \biggl\{ f_x(\cdot) = \frac{1}{\sqrt{h^{d+2}}}
N(x) H_N^{-1}(x) N(x)^T (\nabla K) \biggl(
\frac{x-\cdot}{h} \biggr), x\in R_h \biggr\}.
\end{equation}
Define the empirical process
$(\mathbb{G}_n(f)\dvtx f\in{\cal F})$
where
%
\begin{equation}
\mathbb{G}_n(f) = \frac{1}{\sqrt{n}}\sum
_{i=1}^n \bigl(f(X_i)-\mathbb{E}
\bigl(f(X_i)\bigr) \bigr).
\end{equation}

\begin{thmm}[(Local uncertainty theorem)] \label{LU2}
Assume \textup{(K1)--(K2)}, \textup{(P1)--(P2)}.
Suppose that
$\frac{nh^{d+8}}{\log n}\rightarrow\infty$.
If $h\to0$, then we further assume \textup{(P3)}.
Then for all $x\in R_h$,
when $\|\hat{p}_h-p_h\|^*_{\infty,4}$ is sufficiently small,
\[
\sup_{x\in R_h}\bigl\|\sqrt{nh^{d+2}} \mathbf{d}(x,
\hat{R}_h) - \mathbb {G}_n(f_x)
\bigr\|_{\infty} = O\bigl(\|\hat{p}_h-p_h
\|^*_{\infty,3}\bigr) =O_P \biggl(\sqrt{\frac{\log n}{nh^{d+6}}}
\biggr)
\]
and
$nh^{d+2}\rho^2_n(x) = \operatorname{ Trace}(\Sigma(x))+o(1)$,
where
\[
\Sigma(x) = \operatorname{ Cov} \bigl(N(x) H_N^{-1}(x)
N(x)^T\nabla K(x-X_i) \bigr).
\]
\end{thmm}

We used Theorem~\ref{LUlem2} to convert the rate
$O(\|\hat{p}_h-p_h\|^*_{\infty,3})$ into
$O_P (\sqrt{\frac{\log n}{nh^{d+6}}}  )$ in the first
equality. An intuitive\vspace*{1pt}
explanation for the approximation error rate
$\|\hat{p}_h-p_h\|^*_{\infty,3}$ comes from difference in normal
matrices $N(x)N(x)^T$ and $\hat{N}_n(x)\hat{N}_n(x)$ by claim 6 in
Lemma~\ref{lem::NS}.

\begin{remark}
For a fixed $x$, $\mathbb{G}_n(f_x)$ is a vector and converges to a
mean $0$ multivariate-normal distribution with covariance matrix
$\Sigma(x)$ having rank $d-1$. This theorem also shows the
asymptotic result for the local uncertainty measure
$\rho^2_n(x)$. The matrix $\Sigma(x)$ determines the behavior of
$\rho^2_n(x)$ and depends on three quantities: the normal matrix
$N(x)$, the inverse of subspace Hessian $H_N^{-1}(x) $ and the kernel
function $\nabla K(x-X_i)$.
The normal matrix comes from the fact
that $\mathbf{d}(x,\hat{R}_h)$ is asymptotically in the normal space of
$R_h$ at $x$.
The inverse of subspace Hessian $H_N^{-1}(x)$
plays the same role as the inverse Hessian to a local mode.
We will discuss its properties later.
The last term comes from the kernel density estimator
that depends on the kernel function we use.
\end{remark}

Theorem~\ref{LU2} shows that the uncertainty measure
has a limiting distribution that is similar to KDE for estimating the gradient.
The difference is the matrix $N(x) H_N^{-1}(x) N(x)^T$ whose properties
are given in the following lemma.

\begin{lem}
\label{lem::SH}
Assume \textup{(P1)--(P2)}. Let
\[
W(x) = N(x)^T H_N^{-1}(x)N(x)^T=
N(x)^T \bigl(N(x)^T H(x)N(x) \bigr)^{-1}N(x)^T.
\]
Then:
\begin{longlist}[(1)]
\item[(1)] For any other
$d\times(d-1)$ matrix $N'(x)$ such that $N'(x)^TN'(x) =\mathbf
{I}_{d-1}$ and $N'(x)N'(x)^T= N(x)N(x)^T$,
\[
N(x)^T \bigl(N(x)^T H(x)N(x) \bigr)^{-1}N(x)^T
= N'(x)^T \bigl(N'(x)^T
H(x)N'(x) \bigr)^{-1}N'(x)^T
\]
when $x\in R_h\oplus\delta_0$.
\item[(2)] When $\|x-y\|$ is sufficiently small,
\[
\bigl\|W(x)-W(y)\bigr\|_{\max} \leq A_3 \bigl(\bigl\|q^{(3)}
\bigr\|_{\infty}+\bigl\|q^{(4)}\bigr\| _{\infty}\bigr)^2\|x-y\|
\]
for some constant $A_3$.
%
%
\item[(3)] Assume another density $q$ satisfies \textup{(A1)} and \textup{(P1)}, and let
$W_q(x)$ be the counterpart of $W(x)$ for density $q$.
When $\|p_h-q\|^*_{\infty,3}$ is sufficiently small,
\[
\bigl\|W(x)-W_q(x)\bigr\|_{\max} \leq A_4
\|p_h-q\|^*_{\infty,3}
\]
for some constant $A_4$.
\end{longlist}
\end{lem}

The proof can be found the supplementary material [\citet{chen2014ridgesupp}].
The first result shows that the matrix $N(x)^T H_N^{-1}(x)\times\break N(x)^T$
is the same for any orthonormal matrix $N'(x)$ whose column space
spans the same space.
This shows that $W(x)$ is unaffected if multiplicity of eigenvalues occur.
The second result gives the smoothness for $N(x)^T\times\break  H_N^{-1}(x)N(x)^T$, and
the third result shows stability under small perturbation on the density.

Now we show that the uncertainty measure $\rho^2_n(x)$ can be
estimated by the bootstrap. Given the observed data $\mathbb{X}_n
=\{X_1,\ldots,X_n\}$, we generate the bootstrap sample
$\mathbb{X}^*_n=\{X_1^*,\ldots,X_n^*\}$. We use the bootstrap sample to
construct the bootstrap KDE
%
\begin{equation}
\hat{p}^*_h(x) = \frac{1}{nh^d} \sum
_{i=1}^n K \biggl(\frac
{x-X^*_i}{h} \biggr).
\end{equation}
The bootstrap ridge is
%
\begin{equation}
\hat{R}^*_h = \Ridge\bigl(\hat{p}^*_h\bigr).
\end{equation}
Let
%
\begin{equation}
\hat{\rho}^2_n(x) = \cases{ \mathbb{E} \bigl(d\bigl(x,
\hat{R}^*_h\bigr)^2|X_1,\ldots,
X_n \bigr),& \quad$\mbox {for } x\in\hat{R}_h$,\vspace*{2pt}
\cr
0, &\quad $\mbox{otherwise},$}
\end{equation}
be the bootstrap estimate to the local uncertainty measure.

\begin{thmm}[(Bootstrap consistency)] \label{BT}
Assume \textup{(K1)--(K2)}, \textup{(P1)--(P2).}
For all large $n$ the following is true.
There exists an event
$\mathcal{X}_n$ such that
$\P(\mathcal{X}_n)\geq1-5e^{-nh^{d+8}D_1}$ for some constant $D_1$,
and for
$\mathbb{X}_n\in\mathcal{X}_n$,
when $\|\hat{p}_h-p_h\|^*_{\infty,4}$ is sufficiently small,
for all $x\in R_h$:
\begin{longlist}[(1)]
\item[(1)] The set $\hat{R}_h\cap B(x,\Haus(\hat{R}_h,R_h))\neq
\phi$.
\item[(2)] The estimated ridge satisfies: $\hat{R}_h = \bigcup_{x\in
R_h} (\hat{R}_h\cap B(x,\Haus(\hat{R}_h,R_h)) )$.
\item[(3)]
Suppose that $\frac{nh^{d+8}}{\log n}\rightarrow\infty$.
The estimated local uncertainty measure is consistent in the sense that
for any
$y\in\hat{R}_h\cap B(x,\Haus(\hat{R}_h,R_h))$,
\[
nh^{d+2}\bigl\llvert \hat{\rho}^2_n(y) -
\rho^2_n(x)\bigr\rrvert = O\bigl(\|\hat{p}_h-p_h
\|^*_{\infty,3}\bigr) = O_P \biggl(\sqrt{\frac{\log
n}{nh^{d+6}}}
\biggr).
\]
\end{longlist}
[If we allow $h\rightarrow0$, we need to assume \textup{(P3)}.]
\end{thmm}

Note that we need the above set-based argument because $\rho_n(x)$ and
$\hat{\rho}_n(x)$ are defined on different supports: $\rho_n(x)$ is
defined on
$R_h$ while $\hat{\rho}_n(x)$ is defined on $\hat{R}_h$. This theorem
shows that as $\hat{R}_h$ is approaching $R_h$, the estimated local
uncertainty on $\hat{R}_h$ will converge to the local uncertainty
defined on $R_h$.

\subsection{Gaussian approximation}
\label{sec::GP}
In this section, we derive the limiting distribution of the Hausdorff
distance.
Let $\mathbb{B}$ be a centered, tight Gaussian process defined on
$\mathcal{F}$ with covariance function
%
\begin{equation}
\Cov\bigl(\mathbb{B}(f_1),\mathbb{B}(f_2)\bigr) =
\mathbb {E}\bigl[f_1(X_i),f_2(X_i)
\bigr] - \mathbb{E}\bigl[f_1(X_i)\bigr] \mathbb{E}
\bigl[f_2(X_i)\bigr].
\end{equation}
Such Gaussian processes exists if $\mathcal{F}$ is pre-Gaussian.
The kernel functions and its derivatives of order less than four are
pre-Gaussian by assumption (K2).

\begin{thmm}[(Gaussian approximation)]
\label{thmm::GP}
Assume conditions \textup{(K1)--(K2)}, \textup{(P1)--(P2)} and
that $\frac{nh^{d+8}}{\log n}\rightarrow\infty$. Then there exists
a Gaussian process
$\mathbb{B}$ defined on a function space $\mathcal{F}_h$ [see
equation~\eqref{eq::Ch1}]
such that, when $n$ is sufficiently large,
\[
\sup_{t} \Bigl|\P \bigl(\sqrt{nh^{d+2}}\Haus(
\hat{R}_h,R_h)<t \bigr)- \P \Bigl(\sup
_{f\in\mathcal{F}_h}\bigl|\mathbb{B}(f)\bigr|<t \Bigr) \Bigr| =O \biggl(\frac{\sqrt{\log n}}{(nh^{d+2})^{1/8}}
\biggr).
\]
We can replace $\Haus(\hat{R}_h,R_h)$ with $\dist_{\Pi}(\hat
{R}_h,R_h)$ in the above.
If we allow $h\rightarrow0$, we need to assume \textup{(P3)}.
\end{thmm}

Here we provide an intuitive explanation.
From Theorem~\ref{LU2}, the local uncertainty vector
$\mathbf{d}(x,\hat{R}_h)$ can be approximated by an empirical process.
Recall from equations~\eqref{eq::PR1}, \eqref{eq::PR2}, \eqref
{eq::aHaus}, we have
%
\begin{equation}
\dist_{\Pi}(\hat{R}_h,R_h) =
\sup_{x\in R_h} \bigl\|\mathbf {d}(x,\hat{R}_h)\bigr\|.
\end{equation}
The Hausdorff distance and the quasi-Hausdorff distance will be
the same when the two ridges are close enough; see Lemma~\ref{lem::AH}.
The above argument shows the connection between Hausdorff distance and the
empirical process. The rest of the proof of Theorem~\ref{thmm::GP}
establishes the approximation of the empirical process by the Gaussian
process and applies an anti-concentration argument due to
\citet{CHERNOZHUKOV2013a} to construct the Berry--Esseen type bound.

\begin{remark}
As a referee points out, the Hausdorff distance is usually unstable.
Here we obtain a nice concentration because of assumption (P1)--(P2) along
with the fact that $p_h$ has fourth derivatives.
These conditions ensure the density near ridges is well behaved.
\end{remark}


\subsection{Asymptotic validity of the confidence set}
\label{sec::GPB}

To show our confidence set is consistent, we need to show that
\[
\hat F(t) = \P \bigl(\sqrt{n h^{d+2}}\dist_{\Pi}\bigl(
\hat{R}^*_h,\hat {R}_h\bigr) < t |
\mathbb{X}_n \bigr)
\]
has the same limit as
\[
F(t) = \P \bigl(\sqrt{n h^{d+2}}\dist_{\Pi}(
\hat{R}_h,R_h) < t \bigr).
\]


\begin{thmm}[(Gaussian approximation for bootstrapping)]
\label{thmm::GPB}
Assume conditions \textup{(K1)--(K2)} and \textup{(P1)--(P2)} and that
$\frac{nh^{d+8}}{\log n}\rightarrow\infty$.
For all large $n$ the following is true.
There exists an event
$\mathcal{X}_n$ such that
$\P(\mathcal{X}_n)\geq1-5e^{-nh^{d+8}D_1}$ for some constant $D_1$,
and for
$\mathbb{X}_n\in\mathcal{X}_n$, there exists a Gaussian process
$\mathbb{B}$ defined on a space $\mathcal{F}_h$ [see equation~\eqref
{eq::Ch1}]
such that
\begin{eqnarray*}
&&\sup_t \Bigl| \P \bigl(\sqrt{nh^{d+2}}\Haus\bigl(
\hat{R}^*_h,\hat {R}_h\bigr) < t |
\mathbb{X}_n \bigr) - \P \Bigl(\sup_{f\in\mathcal{F}_h} \bigl|
\mathbb{B}(f)\bigr| < t \Bigr)\Bigr |
\\
&&\qquad= O \biggl(\frac{\sqrt{\log n}}{(nh^{d+2})^{1/8}} \biggr) + O \biggl( \biggl(\frac{\log n}{nh^{d+6}}
\biggr)^{1/6} \biggr).
\end{eqnarray*}
A similar result also holds
when replacing $\Haus(\hat{R}^*_h,\hat{R}_h)$ by $\dist_{\Pi}(\hat
{R}^*_h,\hat{R}_h)$.
Note that if we allow $h\rightarrow0$, we need to assume \textup{(P3)}.
\end{thmm}

The above result, together with Theorem~\ref{thmm::GP}, establishes a
Berry--Esseen result for the bootstrap estimate for the distribution of
$\Haus(\hat{R}_h,R_h)$. Theorem~\ref{thmm::GPB} gives the rate for the
bootstrap case.

\begin{remark}
One might expect the rate to be $O_P (\frac{\sqrt{\log
n}}{(nh^{d+2})^{1/8}} )$
in light of Theorem~\ref{thmm::GP}. The
second term $O_P((\frac{\log n}{nh^{d+6}})^{1/6})$ comes from the
difference in support of the two ridges $R_h,\hat{R}_h$.
The rate is related to the rate estimating the
third derivative of a density, which contributes to the difference
in normal spaces between points of $R_h$ and $\hat{R}_h$.
\end{remark}

We now have the following result on the coverage of the confidence set.

\begin{thmm}
\label{thmm::GP5}
Assume \textup{(K1)--(K2)}, \textup{(P1)--(P2)} and that $\frac{nh^{d+8}}{\log n} \to\infty$.
Let $\hat{t}_{\alpha}=\hat F^{-1}(1-\alpha)$.
Then
\[
\P\bigl(R_h\subset\hat{R}_h\oplus\hat{t}_{\alpha}/
\sqrt{n h^{d+2}}\bigr) \geq1-\alpha+ O \biggl(\frac{\log n}{(nh^{d+2})^{1/8}} \biggr)
+ O \biggl( \biggl(\frac{\log n}{nh^{d+6}} \biggr)^{1/6} \biggr).
\]
If we allow $h\rightarrow0$, we need to assume \textup{(P3)}.
\end{thmm}

This theorem is a direct result of
Theorems \ref{thmm::GP} and \ref{thmm::GPB}, so we omit the proof.
Note that here $\hat{t}_{\alpha}$ differs to the one defined in
Algorithm \ref{Alg::CI}
(and Section~\ref{sec::CI})
by a factor $\sqrt{n h^{d+2}}$.
This is because we rescale $\dist_{\Pi}(\hat{R}^*_h,\hat{R}_h)$
when defining
$\hat{F}(t)$.

\begin{remark}
As a referee points out, one can use
\[
\psi_h = \max_{x\in R_h} \frac{d(x,\hat{R}_h)}{p_h(x)}
\]
as a replacement for $\dist_{\Pi}(\hat{R}_h,R_h)$
and use the bootstrap
to construct a confidence set.
This is a variance-stabilizing version for the original confidence set.
This confidence set is also valid by a simple modification of
Theorems~\ref{thmm::GP}--\ref{thmm::GP5}.
\end{remark}

\section{Examples} \label{sec::example}

We consider two simulation settings: the circle data and the smoothed
box data.
For all simulations, we use a sample
size of $500$. We choose the bandwidth $h$ using Silverman's rule
[\citet{Silverman1986}].

\begin{figure}

\includegraphics{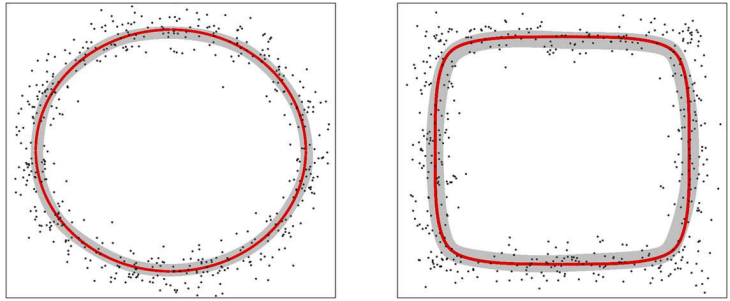}

\caption{$90\%$ confidence sets for the circle data (left) and box
data (right).
The red curve is the smoothed ridge $R_h$, and
the gray regions are confidence sets.}
\label{Fig::CC1}
\end{figure}

The first dataset is the circle data. See Figure~\ref{Fig::CC1}.
We show the true smoothed ridge (red) and the
estimated ridge (blue) along with the $90\%$ confidence sets (gray
regions).

The second dataset is the box data;
see Figure~\ref{Fig::CC1}. Notice that the original box data has corners
that violate condition (P1), but the ridge of the smoothed density $p_h$
obeys (P1).
We show the $90\%$ confidence sets. The box
data has a large angle near its corner, but our confidence
set still has good behavior over these regions.

\section{Proofs} \label{sec::pf}
We prove the main theorems in this section.
The proofs for the lemmas (including those used for proving the main theorems)
are given in the supplementary material;
see \hyperref[suppA]{Supplementary proofs} and \citet{chen2014ridgesupp}.
Before we prove Theorem~\ref{LU2}, we state three useful lemmas.

\begin{lem}
\label{LUlem3}
Let $R$ be the ridge of a density $p$.
For $x\in R$,
let the Hessian at $x$ be $H(x)$ with eigenvectors $[v_1,\ldots, v_d]$
and eigenvalues $0>\lambda_2\geq\cdots\geq\lambda_d$. Consider any
subspace $\mathbb{L}$ spanned by a basis $[e_2,\ldots, e_d]$ with $e_1$
in the normal direction of that subspace. Then a sufficient condition for
$x$ being a local mode of $p$ constrained to
$\mathbb{L}$ is
%
\begin{equation}
\bigl(v_1^Te_1 \bigr)^2 >
\frac{\lambda_1}{\lambda_1-\lambda_2}.
\end{equation}
\end{lem}

The proof can be found in the supplementary material [\citet{chen2014ridgesupp}].

The following lemma is a uniform bound for the KDE.

\begin{lem}[{[\citet{Gine2002}; version of~\citet{Genovese2012a}]}]
\label{LUlem2}
Assume \textup{(K1)--(K2)}
and that
$\log n/n \leq h^d \leq b$ for some $0 < b < 1$.
Then we have
%
\begin{equation}
\Vert\hat{p}_{n}-p\Vert_{\infty, k} = O\bigl(h^2\bigr)
+ O_P \biggl(\sqrt {\frac{\log
n}{nh^{d+2k}}} \biggr)
\end{equation}
for $k=0,\ldots,4$.
In particular, if we consider the smoothed version of density, $p_h$,
for the same kernel function, then we have
\[
\Vert\hat{p}_{n}-p_h\Vert_{\infty, k} =
O_P \biggl(\sqrt{\frac
{\log
n}{nh^{d+2k}}} \biggr)
\]
for $k=0,\ldots,4$.
\end{lem}

\begin{lem}
\label{LUlem4}
Assume \textup{(K1)--(K2)}. Then we have
%
\begin{equation}
\mathbb{E} \bigl( \bigl(\Vert\hat{p}_{n}-p_h
\Vert^*_{\infty,
k} \bigr)^2 \bigr) = O \biggl(\frac{\log n}{nh^{d+2k}}
\biggr)
\end{equation}
for $k=0,\ldots,4$.
\end{lem}

This lemma follows directly from Talagrand's inequality~[\citet{Talagrand1996}],
which proves an exponential concentration inequality for random
variable $\Vert\hat{p}_{n}-p_h\Vert^*_{\infty, k}$.
Thus the second moment is bounded at the specified rate.

In the next proof,
we will frequently use the following theorem that links
the uniform derivative difference to the Hausdorff distance.
%
\begin{thmm}[{[Theorem~6 in~\citet{Genovese2012a}]}] \label{LU0}
Assume condition \textup{(A1)}, \textup{(P1)} for two densities $p_1,p_2$. When
$\Vert p_1-p_2\Vert^*_{\infty, 3}$ is sufficiently small, we have
$\Haus(R_1,R_2) = O(\Vert p_1-p_2\Vert^*_{\infty, 2})$.
\end{thmm}

\begin{pf*}{Proof of Theorem~\ref{LU2}}
Theorem~\ref{LU2} makes two claims: the first claim is an empirical
approximation
\[
\sup_{x\in R_h}\bigl\|\sqrt{nh^{d+2}}\mathbf{d}(x,
\hat{R}_h) - \mathbb {G}_n(f_x)
\bigr\|_{\infty} = O\bigl(\|\hat{p}_h-p_h
\|^*_{\infty,3}\bigr),
\]
and the second claim is the limiting behavior for the uncertainty
measure $\rho^2_n(x)$.
We prove the empirical approximation first and then use it to show the
asymptotic theory for the uncertainty measure.

\textit{Proof for the empirical approximation}.
Let $g_h(x) = \nabla p_h(x)$ and $\hat{g}_n(x) = \nabla\hat{p}_h(x)$,
and define $N(x), \hat{N}_n(x)$ to be the normal space at $x\in R_h$ and
$x\in\hat{R}_h$, respectively. Note that when
$\|\hat{p}_h-p_h\|^*_{\infty,2}$ is sufficiently small, we have
(P1) for~$\hat{p}_h$. This implies that
$N(x),\hat{N}_n(x)$ can be
defined (but they are not necessarily unique)
for points near $R_h,\hat{R}_h$
by claim 3 of Lemma~\ref{lem::NS}.
Condition (P3) ensures that the
constants in (P1) and the reach of $p_h$ have positive lower bound as
$h\rightarrow0$ for $p_h$.

By (P2) and Lemma~\ref{LUlem3}, the ridges are the local modes in the
subspace $N(x)$. Note that despite the fact that $N(x)$ may not be
unique, the column space of $N(x)$ is unique by claim 5 in
Lemma~\ref{lem::NS}. Hence we have
%
\begin{equation}
N(x)^T g_h(x)=0,\qquad \hat{N}_n(z)
\hat{g}_n(z)=0 \label{eq::LU2pf1}
\end{equation}
for all $x\in R_h$ and $z\in\hat{R}_h$.
This shows that ridges are generalized local modes with respect to
their local normal coordinate.

Let $\tilde{x} = \pi_{\hat{R}_h}(x)\in\hat{R}_h$.
When $\|\tilde{x}-x\|$ is smaller than the reach of $\hat{R}_h$, the
projection $\tilde{x}$ is unique. By claim 7 in Lemma~\ref{lem::NS}
and the
fact that $\Haus(\hat{R}_h,R_h) = O(\|\hat{p}_h-p_h\|^*_{\infty,2})$
from Theorem~\ref{LU0}, the reach of $\hat{R}_h$ and the reach of
$R_h$ will be close once $\|\hat{p}_h-p_h\|^*_{\infty,4}$ is
sufficiently small. Accordingly, $\mathbf{d}(x,\hat{R}_h) = x-\tilde{x}$
is
unique
once $\|x-\tilde{x}\|\leq\Haus(\hat{R}_h,R_h) =
O(\|\hat{p}_h-p_h\|^*_{\infty,2})$ is sufficiently small.
This leads to
%
\begin{eqnarray}
 \hat{g}_n(\tilde{x})-g_h(x)
& =&\hat{g}_n(\tilde{x})-g_h(\tilde{x})+g_h(
\tilde{x})-g_h(x)\nonumber
\\
&\leq& O\bigl(\|\hat{p}_h-p_h\|^*_{\infty,1}+
\Haus(\hat{R}_h,R_h)\bigr)
\\
& \leq &O\bigl(\|\hat{p}_h-p_h\|^*_{\infty,2}\bigr)
\qquad\mbox{by Theorem~\ref{LU0}.} \nonumber
\end{eqnarray}
We use the fact that $g_h(x)$ has bounded derivatives
from (K1). Accordingly, $\hat{g}_n(\tilde{x})$
converges
to $g_h(x)$.
Hence, when
$\|\hat{p}_h-p_h\|^*_{\infty,2}$ is sufficiently small,
$N(x)^T\hat{g}_n(\tilde{x})=0$ by Lemma~\ref{LUlem3}. Since
$N(x)^T\hat{g}_n(\tilde{x})=0$,
%
\begin{equation}
N(x)^T\hat{g}_n(\tilde{x})=0=N(x)^T\bigl[
\hat{g}_n(\tilde{x})-\hat {g}_n(x)+\hat{g}_n(x)-g_h(x)
\bigr],
\end{equation}
which leads to
%
\begin{eqnarray}\label{eq::LU2pf2}
 N(x)^T\bigl[\hat{g}_n(
\tilde{x})-\hat{g}_n(x)\bigr]&=&-N(x)^T\bigl[\hat
{g}_n(x)-g_h(x)\bigr]
\nonumber
\\[-8pt]
\\[-8pt]
\nonumber
&=&-N(x)^T \bigl[\hat{g}_n(x)-\mathbb{E} \bigl(
\hat{g}_n(x) \bigr)\bigr].
\end{eqnarray}
We used
$g_h(x) = \mathbb{E} (\hat{g}_n(x) )$
in the last
equality. Since $\|\tilde{x}-x\|$ is small due to Theorem~\ref
{LU0}, and
$\|\hat{p}_h-p\|^*_{\infty}$ is small, we use Taylor's theorem
for the first term which yields
%
\begin{eqnarray}\label{eq::LU2pf3}
&&N(x)^T \bigl[\hat{g}_n(
\tilde{x})-\hat{g}_n(x)\bigr]\nonumber
\\
& &\qquad= N(x)^T \int_{0}^1
\hat{H}_n\bigl(x+(\tilde{x}-x)t\bigr)\,dt (\tilde{x}-x)
\nonumber
\\[-8pt]
\\[-8pt]
\nonumber
&& \qquad= N(x)^T H(x) \bigl(1+O\bigl(\|\hat{p}_h-p_h
\|^*_{\infty,2}\bigr)+O\bigl(\|\tilde {x}-x\|\bigr) \bigr) (\tilde{x}-x)
\\
&&\qquad= N(x)^T H(x) (\tilde{x}-x) \bigl(1+O\bigl(\|\hat
{p}_h-p_h\|^*_{\infty
,2}\bigr) \bigr).\nonumber
\end{eqnarray}
We use the fact that
\[
\int_{0}^1 \hat{H}_n\bigl(x+(
\tilde{x}-x)t\bigr)\,dt = H(x) \bigl(1+O\bigl(\|\hat{p}_h-p_h
\|^*_{\infty,2}\bigr)+O\bigl(\|\tilde {x}-x\|\bigr) \bigr)
\]
in the second equality and apply Theorem~\ref{LU0} to absorb $O(\|
\tilde{x}-x\|)$
into the other term.
By claims 5, 6
in Lemma~\ref{lem::NS} and the fact that
the line segment joining $\tilde x$ and $x$
is contained in
$R_h\oplus\delta_0$ by (P1), we have
%
\begin{equation}
\bigl\|\hat{N}_n(\tilde{x})\hat{N}_n(\tilde{x})^T-N(x)N(x)^T
\bigr\|_{\max
} = O\bigl(\|\hat{p}_h-p_h
\|^*_{\infty,3}\bigr). \label{eq::LU2pf4}
\end{equation}

Now $\tilde{x}-x= \hat{N}_n(\tilde{x})\hat{N}_n(\tilde{x})^T
(\tilde{x}-x)$. Combining this with
equations~\eqref{eq::LU2pf2}, \eqref{eq::LU2pf3} and \eqref{eq::LU2pf4}
we obtain
%
\begin{eqnarray}
&& -N(x)^T \bigl[\hat{g}_n(x)-
\mathbb{E}\bigl(\hat{g}_n(x)\bigr)\bigr]\nonumber
\\
&&\qquad= N(x)^T\bigl[\hat{g}_n(\tilde{x})-
\hat{g}_n(x)\bigr]\nonumber
\\
&&\qquad = N(x)^T H(x) (\tilde{x}-x) \bigl(1+O\bigl(\|\hat{p}_h-p_h
\|^*_{\infty
,2}\bigr)\bigr)
\nonumber
\\[-8pt]
\\[-8pt]
\nonumber
&&\qquad = N(x)^T H(x)\hat{N}_n(\tilde{x})\hat{N}_n(
\tilde{x})^T (\tilde {x}-x) \bigl(1+O\bigl(\|\hat{p}_h-p_h
\|^*_{\infty,2}\bigr)\bigr)
\\
&&\qquad = N(x)^T H(x)N(x)N(x)^T (\tilde{x}-x) \bigl(1+O\bigl(
\|\hat {p}_h-p_h\| ^*_{\infty,3}\bigr)\bigr)\nonumber
\\
&&\qquad = H_N(x) N(x)^T (\tilde{x}-x) \bigl(1+O\bigl(\|
\hat{p}_h-p_h\|^*_{\infty,3}\bigr)\bigr),\nonumber
\end{eqnarray}
where
%
\begin{equation}
H_N(x) = N(x)^T H(x)N(x).
\end{equation}
In the fourth equality, we used~\eqref{eq::LU2pf4}.
Multiplying the matrix $H_N(x)$ to the left of both sides
and moving $O(\|\hat{p}_h-p_h\|^*_{\infty,3})$ to the other side,
%
\begin{eqnarray}\label{eq::LU2pf5}
&& N(x)^T (\tilde{x}-x)
\nonumber
\\[-8pt]
\\[-8pt]
\nonumber
&&\qquad= -H_N(x)^{-1}N(x)^T \bigl[
\hat{g}_n(x)-\mathbb{E}\bigl(\hat{g}_n(x)\bigr)\bigr]
\bigl(1+O\bigl(\|\hat{p}_h-p_h\|^*_{\infty,3}\bigr)
\bigr).
\end{eqnarray}

We multiply by $N(x)$ and use \eqref{eq::LU2pf4} again to obtain
%
\begin{eqnarray}\label{eq::LU2pf5-1}
 N(x)N(x)^T (\tilde{x}-x) &=&
\hat{N}_n(\tilde{x})\hat{N}_n(\tilde {x})^T(
\tilde{x}-x) \bigl(1+O\bigl(\|\hat{p}_h-p_h
\|^*_{\infty,2}\bigr)\bigr)
\nonumber
\\[-8pt]
\\[-8pt]
\nonumber
& =& (\tilde{x}-x) \bigl(1+O\bigl(\|\hat{p}_h-p_h
\|^*_{\infty,3}\bigr)\bigr).
\end{eqnarray}
Let $W_2(x) = N(x)H_N(x)^{-1}N(x)^T$, and define
$\mathbf{d}(x,\hat{R}_h) = \tilde{x}-x$.
Combining
\eqref{eq::LU2pf5} and \eqref{eq::LU2pf5-1},
%
\begin{equation}
\mathbf{d}(x,\hat{R}_h) = W_2(x)\bigl[
\hat{g}_n(x)-\mathbb{E}\bigl(\hat {g}_n(x)\bigr)\bigr]
\bigl(1+O\bigl(\|\hat{p}_h-p_h\|^*_{\infty,3}\bigr)
\bigr). \label{eq::LU2pf6}
\end{equation}
Notice that the KDE can be expressed in terms of the empirical process via
%
\begin{eqnarray}\label{eq::LU2pf7}
&& \hat{g}_n(x) - \mathbb{E}\bigl(
\hat{g}_n(x)\bigr) 
 \biggl(\frac{x-X_i}{h} \biggr) -
\mathbb{E} \biggl(\frac
{1}{h^{d+2}} (\nabla K ) \biggl(\frac{x-X_i}{h} \biggr)
\biggr)
\nonumber
\\[-8pt]
\\[-8pt]
\nonumber
&&\qquad=\frac{1}{\sqrt{n}}\mathbb{G}_n(\tau_x),
\end{eqnarray}
where $y\mapsto\tau_x(y) = \frac{1}{h^{d+1}}(\nabla K)(\frac
{x-y}{h}) $.
From equation~\eqref{eq::LU2eq1},
%
\begin{equation}
f_x(y) = \sqrt{h^{d+2}}W_2(x)
\tau_x(y) \label{eq::LU2pf8}
\end{equation}
for all $x\in R_h$.
Hence, multiplying \eqref{eq::LU2pf6}
by $\sqrt{nh^{d+2}}$
and using \eqref{eq::LU2pf7} and \eqref{eq::LU2pf8},
%
\begin{equation}
\sqrt{nh^{d+2}}\mathbf{d}(x,\hat{R}_h) -
\mathbb{G}_n(f_x) = O\bigl(\|\hat{p}_h-p_h
\|^*_{\infty,3}\bigr), \label{eq::LU2pf9}
\end{equation}
for each $x\in R_h$.
Note that the bound $O(\|\hat{p}_h-p_h\|^*_{\infty,3})$ is
independent of $x$
and the above construction is
valid
for all $x \in R_h$.
Hence
\[
\sup_{x\in R_h}\bigl\|\sqrt{nh^{d+2}}\mathbf{d}(x,
\hat{R}_h) - \mathbb {G}_n(f_x)
\bigr\|_{\infty} = O\bigl(\|\hat{p}_h-p_h
\|^*_{\infty,3}\bigr).
\]
This proves the approximation for $\mathbf{d}(x,\hat{R}_h)$.

\textit{Proof for the uncertainty measures}.
We first prove that the local uncertainty measure $nh^{d+2}\rho_n(x)$
converges to
$\mathbb{E} (\|\mathbb{G}_n(f_x)\|^2 )$. Then we show
the limiting
behavior for $\|\mathbb{G}_n(f_x)\|$. We have
%
\begin{eqnarray}\label{eq::LU2pf10}
&&\bigl|nh^{d+2}\rho_n(x) - \mathbb{E}
\bigl(\bigl\|\mathbb {G}_n(f_x)\bigr\| ^2 \bigr) \bigr|\nonumber
\\
&&\hspace*{76pt}\quad = \bigl\llvert \mathbb{E} \bigl(nh^{d+2}d^2(x,
\hat{R}_h)-\bigl\|\mathbb {G}_n(f_x)
\bigr\|^2 \bigr)\bigr\rrvert\nonumber
\\
&&\hspace*{41pt}\mbox{(Jensen's)}\leq\mathbb{E}\bigl\llvert
nh^{d+2}d^2(x,\hat {R}_h)-\bigl\|
\mathbb{G}_n(f_x)\bigr\|^2\bigr\rrvert
\nonumber
\\[-8pt]
\\[-8pt]
\nonumber
&&\hspace*{76pt}\quad=\mathbb{E}\bigl\llvert nh^{d+2}\bigl\|\mathbf{d}(x,\hat{R}_h)
\bigr\|^2-\bigl\|\mathbb {G}_n(f_x)\bigr\|^2
\bigr\rrvert
\\
&&\hspace*{76pt}\quad= \mathbb{E} \bigl\llvert (D_n-G_n )^T
(D_n+G_n )\bigr\rrvert\nonumber
\\
&&\mbox{(Cauchy--Schwarz)}\leq\sqrt{\mathbb{E} \bigl( \|D_n-G_n
\| ^2 \bigr)\mathbb{E} \bigl( \|D_n+G_n
\|^2 \bigr)},\nonumber
\end{eqnarray}
where $D_n = \sqrt{nh^{d+2}}\mathbf{d}(x,\hat{R}_h)\in\mathbb
{R}^d$ and
$G_n = \mathbb{G}_n(f_x)\in\mathbb{R}^d$.

Now by \eqref{eq::LU2pf9} and Lemma~\ref{LUlem4},
%
\begin{equation}
\mathbb{E} \bigl( \|D_n-G_n\|^2 \bigr) = O
\biggl(\frac
{1}{nh^{d+6}} \biggr). \label{eq::LU2pf11}
\end{equation}
Note $D_n+G_n \leq2G_n + (D_n-G_n)$, which implies
$\|D_n+G_n\|^2 \leq(\|2G_n\|+\|D_n-G_n\|)^2$.
Taking expectation on both sides and using the fact that
$\sqrt{\mathbb{E}(A^2)}$ is $\cL^2$ norm for the random variable $A$,
%
\begin{eqnarray}
\mathbb{E} \bigl( \|D_n+G_n
\|^2 \bigr) &\leq&\mathbb{E} \bigl(\bigl(\|2G_n\|+
\|D_n-G_n\|\bigr)^2 \bigr)
\nonumber
\\[-8pt]
\\[-8pt]
\nonumber
&\leq& \bigl(\sqrt{\mathbb{E}\bigl(\|2G_n\|^2\bigr)} +
\sqrt{\mathbb {E}\bigl(\|D_n-G_n\|^2\bigr)}
\bigr)^2.
\end{eqnarray}
Again by \eqref{eq::LU2pf9} and Lemma~\ref{LUlem4},
%
\begin{equation}
\mathbb{E} \bigl( \|D_n+G_n\|^2 \bigr) = 2
\mathbb{E}\bigl(\|G_n\| ^2\bigr)+O \biggl(\sqrt{
\frac{1}{nh^{d+6}}} \biggr). \label{eq::LU2pf11-1}
\end{equation}

Here we derive $\mathbb{E}(\|G_n\|^2)$. Recall that
$G_n = \mathbb{G}_n(f_x)$,
where
\[
f_x(\cdot) = \frac{1}{\sqrt{h^{d+2}}} N(x) H_N^{-1}(x)
N(x)^T (\nabla K) \biggl(\frac{x-\cdot}{h} \biggr)
\]
for each $x\in R_h$ by \eqref{eq::LU2eq1}.
Note that $\|G_n\|^2 = G_n^T G_n$ and $\mathbb{E} (G_n
)=0$. Hence
%
\begin{eqnarray}\label{eq::LU2pf12}
 \mathbb{E} \bigl(\|G_n\|^2
\bigr) &= &\mathbb{E} \bigl(G_n^T G_n \bigr)\nonumber
\\
& =& \mathbb{E} \bigl( \bigl(G_n-\mathbb{E} (G_n )
\bigr)^T \bigl(G_n-\mathbb{E} (G_n ) \bigr)
\bigr)
\nonumber
\\[-8pt]
\\[-8pt]
\nonumber
& =& \operatorname{Trace} \bigl(\Cov (G_n ) \bigr)
\\
& = &\operatorname{Trace} \bigl(\Sigma(x) \bigr), \nonumber
\end{eqnarray}
where
\[
\Sigma(x) = \Cov (G_n ) = \Cov \biggl(N(x) H_N^{-1}(x)
N(x)^T (\nabla K) \biggl(\frac{x-X_i}{h} \biggr) \biggr)
\]
is bounded. Thus by \eqref{eq::LU2pf10}--\eqref{eq::LU2pf12} we
conclude that
\begin{eqnarray*}
&&\bigl|nh^{d+2}\rho_n(x) - \mathbb{E} \bigl(\bigl\|\mathbb
{G}_n(f_x)\bigr\| ^2 \bigr) \bigr|
\\
&&\qquad\leq\sqrt{\mathbb{E} \bigl( \|D_n-G_n\|^2
\bigr)\mathbb {E} \bigl( \|D_n+G_n\|^2 \bigr)}
\\
&&\qquad = O \biggl(\sqrt{\frac{1}{nh^{d+6}}} \biggr).
\end{eqnarray*}

Thus the uncertainty measure $\rho_n(x)$ can be approximated by
$\mathbb{E} (\|\mathbb{G}_n(f_x)\|^2 )=\mathbb
{E} (\|G_n\|^2
 )$. Now by \eqref{eq::LU2pf12}, the result follows.
\end{pf*}

Before we prove the bootstrap result, we need the following lemma.

\begin{lem}
\label{lem::PB}
Let $p_h$ be the smoothed density and
$R_h$ be the associated ridges.
Let $\hat{p}_h$ be
the KDE based on the observed data
$\mathbb{X}_n=\{X_1,\ldots,X_n\}$ and
$\hat{R}_h$ be the estimated ridge.
Consider these two conditions:
\begin{longlist}[(T1)]
\item[(T1)] \textup{(P1)--(P2)} holds for $\hat{p}_h$.
\item[(T2)] $\|\hat{p}_h-p_h\|^*_{\infty,4}< s_0$ for a small
constant $s_0$.
\end{longlist}
Let $\mathcal{X}_n = \{\mathbb{X}_n\dvtx\mathrm{(T1)},\mathrm{(T2)}\mbox{ holds}\}$.
Then, when $n$ is sufficiently large,
\[
\P(\mathcal{X}_n) \geq1-5e^{-nh^{d+8}D_1},
\]
for some constant $D_1$.
\end{lem}

The proof can be found in the supplementary material [\citet{chen2014ridgesupp}].

\begin{pf*}{Proof of Theorem~\ref{BT}}
To prove the bootstrap result, we use a technique of
\citet{romano1988bootstrapping} by first considering a sequence of nonrandom
distributions $\{Q_m\dvtx m=1,\ldots\}$. In the last step,
we replace $Q_m$ by the empirical distribution $\P_n$.

Let $q_m$ be the density of the
smoothed distribution $Q_m\star K_h$ where $K_h(x) =
\frac{1}{h^d}K(\frac{x}{h})$ is the kernel\vspace*{1pt} function used in the KDE
and $\star$ is the convolution operator.
If we replace $Q_m$ with the sample distribution $\P$, the smoothed
distribution has density $p_h$.
If we replace $Q_m$ with the empirical
distribution $\hat\P_n$, we obtain the KDE $\hat{p}_h$.

We assume that each smoothed density $q_m$ satisfies conditions
(P1)--(P2) and $\|q_m-p_h\|^*_{\infty,4}\rightarrow0$, and
$\|q_m-p_h\|^*_{\infty,4}$ is sufficiently small for all $m$.
Let $R(q_m)= \Ridge(q_n)$.
Let $\mathbb{Y}_{m,n} =\{ Y_{m,1},\ldots,Y_{m,n}\}$
where
$Y_{m,1},\ldots,Y_{m,n} \sim Q_m$.
Let $\hat{q}_{m,n}$
be the KDE based on $\mathbb{Y}_{m,n}$, and let
$\hat{R}_h(q_m)=\Ridge(\hat{q}_{m,n})$.

Let
$\rho^2_{m,n}(x)$ for $x\in R(q_m)$
be the local uncertainty measure.
When $\|q_m-p_h\|^*_{\infty,3}$ is sufficiently small, we can
apply Theorem~\ref{LU2} to $R(q_m)$ so that
\[
\rho^2_{m,n}(x) = \frac{1}{nh^{d+2}}\operatorname{Trace}
\Sigma(x;q_m) +o \biggl(\frac{1}{nh^{d+2}} \biggr),
\]
where
\[
\Sigma(x;q_m) = \operatorname{Cov}\bigl[N_{q_m}(x)
H_N(x;q_m)^{-1} N_{q_m}(x)^T
\nabla K(x-Y_{m,i})\bigr]
\]
for $x\in R(q_m)$. Note that although we do not assume (P3) for $q_m$,
Theorem~\ref{LU2} is still valid once the gap constants in (P1) have
positive lower bound. In this case, because
$\|\hat{p}_h-p_h\|^*_{\infty,3}$ is sufficiently small and we
assume (P3) for $p_h$, the gap constants have a lower bound for $q_m$ as
$q_m$ approaching $p_h$.

Now we proceed with the proof.
Claims 1 and 2 are trivially true by the definition of Hausdorff distance.
Now we prove claim 3.
When $\|q_m-p_h\|^*_{\infty,3}$ is sufficiently small,
\[
\Haus\bigl(R(q_m), R_h\bigr) = O\bigl(
\|q_m-p_h\|^*_{\infty,2}\bigr).
\]
For any point $x\in R_h$, and any $y\in R_h \cap B(x,\Haus(R(q_m),R_h))$,
\[
\|y-x\|\leq\Haus\bigl(R(q_m), R_h\bigr).
\]

Since $y$ is on $R(q_m)$, the local uncertainty $\rho^2_{m,n}(y)$ is
well defined.
Then
\begin{eqnarray*}\bigl\|\rho^2_{m,n}(y)-
\rho^2_n(x)\bigr\| & = &\frac{1}{nh^{d+2}}\operatorname{Trace} \bigl(
\Sigma(y;q_m)-\Sigma(x)\bigr)+o \biggl(\frac{1}{nh^{d+2}} \biggr)
\\
&\leq&\frac{d}{nh^{d+2}} \bigl\|\Sigma(y;q_m)-\Sigma(x)
\bigr\|_{\max
}+o \biggl(\frac{1}{nh^{d+2}} \biggr).
\end{eqnarray*}
Since the terms in $\Sigma(x)$ involve only the derivatives of
the smoothed density up to the third order and since
$\Vert y-x\Vert\leq\Haus(R(q_m), R_h)$, we conclude that
\begin{eqnarray*}\bigl\|\Sigma(y;q_m)-\Sigma(x)
\bigr\|_{\max} &=& O\bigl(\Haus\bigl(R(q_m), R_h
\bigr)+\|q_m-p_h\|^*_{\infty,3}\bigr)
\\
& =& O\bigl(\|q_m-p_h\|^*_{\infty,3}\bigr).
\end{eqnarray*}
The $O(\|q_m-p_h\|^*_{\infty,3})$
term does not depend on $x$ so that this can be taken uniformly for all
$x\in R_h$. This proves claim 3.

For the bootstrap case, we replace $Q_m$ by
$\hat\P_n$. Thus
$q_m$ is replaced by $\hat{p}_h$
so that we obtain the result.
Notice that we
require that $\hat{p}_h$ satisfies (P1)--(P2) and
that $\|\hat{p}_h-p_h\|^*_{\infty,4}$ be sufficient small.
Recall that $\mathcal{X}_n$ is the collection of $\mathbb{X}_n$ such that
$\hat{p}_h$ satisfies conditions (P1)--(P2) and
$\|\hat{p}_h-p_h\|^*_{\infty,4}$ is sufficiently small.
Applying
Lemma~\ref{lem::PB} we conclude that
\[
\P(\mathcal{X}_n) \geq1-5e^{-nh^{d+8}D_1}
\]
for some constant $D_1$.
\end{pf*}

Before we prove the Gaussian approximation,
we need the following lemma that links the quasi-Hausdorff distance
to the Hausdorff distance.

\begin{lem}
\label{lem::AH}
Let $R_1,R_2$ be two
closed, nonself-intersecting curves with positive reach.
If
\[
\Haus(R_1,R_2)< (2-\sqrt{2})\min \bigl\{\reach
(R_1),\reach(R_2) \bigr\},
\]
then
%
\begin{equation}
\dist_{\Pi}(R_2,R_1)= \dist_{\Pi}(R_1,R_2)
= \Haus(R_1,R_2).
\end{equation}
\end{lem}

The proof can be found in the supplementary material
[\citet{chen2014ridgesupp}].

\begin{pf*}{Proof of Theorem~\ref{thmm::GP}}
Our proof consists of three steps. The first step establishes a
coupling between
the Hausdorff distance $\Haus(\hat{R}_h,R_h)$ and the supremum of an
empirical process. The second step shows that the distribution
of the maxima of the empirical process can be approximated by the
maxima of a Gaussian process. The last step uses
anti-concentration to bound the distributions between
$\Haus(\hat{R}_h,R_h)$ and the maxima of a Gaussian process.

\textit{Step \textup{1}---Empirical process approximation.}

Recall that $\mathbb{G}_n$ is the empirical process defined by
%
\begin{eqnarray}
 \mathbb{G}_n(f_k) &=&
\frac{1}{\sqrt{n}}\bigl(f_k(X_i)-\mathbb {E}
\bigl(f_k(X_i)\bigr)\bigr),
\nonumber
\\[-8pt]
\\[-8pt]
\nonumber
\operatorname{Cov} \bigl(\mathbb{G}_n(f_1),
\mathbb{G}_n(f_2)\bigr) & =& \mathbb {E}
\bigl(f_1(X_1)f_2(X_1)\bigr)
\end{eqnarray}
for any two functions $f_1,f_2$.
We also recall the function $f_x(y)$ in~\eqref{eq::LU2eq1},
%
\begin{equation}
y\mapsto f_x(y) = \frac{1}{\sqrt{h^{d+2}}} N(x) H_N^{-1}(x)
N(x)^T (\nabla K) \biggl(\frac{x-y}{h} \biggr),\qquad x\in
R_h. \label{eq::LU2eq1r}
\end{equation}
Note that $f_x(y)\in\mathbb{R}^d$ is a vector.
Let
%
\begin{equation}
 \mathcal{F}_{h} = \bigl\{
\mathbf{w}^T f_x(y)\dvtx\mathbf{w}\in
\mathbb{R}^{d}, \|\mathbf {w}\|=1, f_x(y) \mbox{ defined
in \eqref{eq::LU2eq1r}},x\in R_h \bigr\}.
\label{eq::Ch1}
\end{equation}
By Theorem~\ref{LU2},
\[
\sup_{x\in R_h}\bigl\|\sqrt{nh^{d+2}}\mathbf{d}(x,
\hat{R}_h) - \mathbb {G}_n(f_x)
\bigr\|_{\infty} = O\bigl(\|\hat{p}_h-p_h
\|^*_{\infty,3}\bigr).
\]
Since the $\cL^2$ norm is bounded by $d$ times the infinity norm for a vector,
\begin{eqnarray*} \sup_{x\in R_h}\bigl\llvert
\sqrt{nh^{d+2}}d(x,\hat{R}_h) - \bigl\|\mathbb
{G}_n(f_x)\bigr\|\bigr\rrvert & =& \sup_{x\in R_h}\bigl\llvert \sqrt{nh^{d+2}}\bigl\|\mathbf{d}(x,\hat {R}_h)\bigr\| - \bigl\|\mathbb{G}_n(f_x)\bigr\|\bigr\rrvert
\\
&= & O\bigl(\|\hat{p}_h-p_h\|^*_{\infty,3}\bigr).
\end{eqnarray*}
For any vector $v\in\mathbb{R}^{d}$,
$\|v\| = \sup_{\|\mathbf{w}\| =1} \mathbf{w}^Tv$
where $\mathbf{w}\in\mathbb{R}^{d}$.
Hence
\[
\Bigl\llvert \sup_{x\in R_h}\sqrt{nh^{d+2}}d(x,
\hat{R}_h) - \sup_{x
\in R_h, \|\mathbf{w}\|=1}\mathbf{w}^T
\mathbb{G}_n(f_x)\Bigr\rrvert = O\bigl(\|
\hat{p}_h-p_h\|^*_{\infty,3}\bigr).
\]

Define
$\|\mathbb{G}_n\|_{\mathcal{F}} = \sup_{f\in\mathcal
{F}}\|\mathbb{G}_n(f)\|$.
Recall that the asymptotic
Hausdorff distance is $\dist_{\Pi}(A,B) = \sup_{x\in B} d(x,A)$.
Then
%
\begin{equation}
\bigl\llvert \sqrt{nh^{d+2}} \dist_{\Pi}(\hat{R}_h,R_h)-
\|\mathbb {G}_n\|_{\mathcal{F}_h}\bigr\rrvert = O\bigl(\|
\hat{p}_h-p_h\|^*_{\infty,3}\bigr). \label{eq::aHaus2}
\end{equation}
This shows that the quasi-Hausdorff distance can be approximated by the supremum
of an empirical process over the functional space $\mathcal{F}_h$.

When $\|\hat{p}_h-p_h\|^*_{\infty,4}$ is sufficiently small, the
reach of $\hat{R}_h$ is close to the reach of $R_h$
by claim 7 of Lemma~\ref{lem::NS},
and the
Hausdorff distance is much smaller than the reach. By
Lemma~\ref{lem::AH}, the quasi-Hausdorff distance is the same as
the Hausdorff distance so that
%
\begin{equation}
\bigl\llvert \sqrt{nh^{d+2}} \Haus(\hat{R}_h,R_h)-
\|\mathbb {G}_n\| _{\mathcal{F}_h}\bigr\rrvert = O\bigl(\|
\hat{p}_h-p_h\|^*_{\infty,3}\bigr). \label{eq::AH1}
\end{equation}
Equation \eqref{eq::AH1} is the coupling between Hausdorff distance
and the supremum of an empirical process and is the main result for step
1. Note that a sufficient condition for
$\|\hat{p}_h-p_h\|^*_{\infty,4}$ being small is that
$\frac{nh^{d+8}}{\log n}\rightarrow\infty$. This is the bandwidth
condition we require.

\textit{Step \textup{2}---Gaussian approximation.}

In this step, we use a theorem of~\citet{CHERNOZHUKOV2013a}
to show that the supremum of the empirical process can be approximated
by the supremum of a centered, tight Gaussian process $\mathbb{B}$
defined on $\mathcal{F}_h$
with covariance function
%
\begin{equation}
\Cov\bigl(\mathbb{B}(f_1),\mathbb{B}(f_2)\bigr) =
\mathbb {E}\bigl[f_1(X_i)f_2(X_i)
\bigr] - \mathbb{E}\bigl[f_1(X_i)\bigr]\mathbb{E}
\bigl[f_2(X_i)\bigr]
\end{equation}
for $f_1,f_2\in\mathcal{F}_h$.
We first recall the theorem of~\citet{CHERNOZHUKOV2013a}.

\begin{thmm}[{[Theorem~3.1 in~\citet{CHERNOZHUKOV2013a}]}]
\label{thmm::GPA1}
Let $\mathcal{G}$ be a collection of functions that is a VC-type class
[see condition \textup{(K2)}] with a constant envelope function $b$. Let $\sigma
^2$ be a constant such that $\sup_{g\in\mathcal{G}}\mathbb
{E}[g(X_i)^2]\leq\sigma^2\leq b^2$. Let $\mathbb{B}$ be a centered,
tight Gaussian process defined on $\mathcal{G}$ with covariance function
%
\begin{equation}
\Cov\bigl(\mathbb{B}(g_1),\mathbb{B}(g_2)\bigr) =
\mathbb {E}\bigl[g_1(X_i)g_2(X_i)
\bigr] - \mathbb{E}\bigl[g_1(X_i)\bigr]\mathbb{E}
\bigl[g_2(X_i)\bigr],
\end{equation}
where $g_1,g_2\in\mathcal{G}$. Then for any $\gamma\in(0,1)$ as $n$
is sufficiently large,
there exists a random variable $\mathbf{B}\stackrel{d}{=}\|\mathbb
{B}\|_{\mathcal{G}}$ such that
%
\begin{equation}
\P \biggl( \bigl|\|\mathbb{G}_n\|_{\mathcal{G}} - \mathbf{B}
\bigr|>A_1\frac{b^{1/3}\sigma^{2/3}\log^{2/3} n}{\gamma^{1/3}
n^{1/6}} \biggr)\leq A_2 \gamma,
\end{equation}
where $A_1,A_2$ are two universal constants.
\end{thmm}

Now we show that $\mathcal{G}$ in Theorem~\ref{thmm::GPA1} can be
linked to $\mathcal{F}_{h}$ with a proper scaling.
From condition (K2),
the collection
\[
\biggl\{t \to \biggl(\frac{\partial}{\partial x_i} K \biggr) \biggl(\frac{x-t}{h}
\biggr)\dvtx x\in\R^d, i=1,\ldots, d \biggr\}
\]
is a VC-type pre-Gaussian class with a constant envelope $b_0$.
Recall equation~\eqref{eq::LU2eq1r}:
\[
f_x(y) = \frac{1}{\sqrt{h^{d+2}}} N(x) H_N^{-1}(x)
N(x)^T (\nabla K ) \biggl(\frac{x-y}{h} \biggr),\qquad x\in
R_h.
\]
This function will not be uniformly bounded as $h\rightarrow0$, so we consider
%
\begin{eqnarray}\label{eq::GP1-1}
g_x(y) &=& \sqrt{h^{d+2}}
f_x(y)
\nonumber
\\[-8pt]
\\[-8pt]
\nonumber
& =&N(x) H_N^{-1}(x) N(x)^T (\nabla K )
\biggl(\frac
{x-y}{h} \biggr),\qquad x\in R_h.
\end{eqnarray}
Note that each element of the vector $g_x(y)$
is uniformly bounded. This is because $N(x) H_N^{-1}(x) N(x)^T \leq
c_1<\infty$
for some universal constant since $N(x)$ is generated by the
derivatives of $p_h$ with order less than four and by (K1) is uniformly bounded.

Now we define
%
\begin{eqnarray}\label{eq::GP1-2}
 \mathcal{G}_h &=& \bigl\{
\mathbf{w}^T g_x(y)\dvtx\mathbf{w}\in
\mathbb{R}^{d}, \|\mathbf {w}\|=1,x\in R_h \bigr\}
\nonumber
\\[-8pt]
\\[-8pt]
\nonumber
& = &\bigl\{\sqrt{h^{d+2}} f\dvtx f\in\mathcal{F}_{h}\bigr
\}.
\end{eqnarray}
Since $\|\mathbf{w}\|=1$ and $N(x) H_N^{-1}(x) N(x)^T \leq c_1$ and
$b_0$ is a constant envelope for the partial derivatives of kernel
functions, $b_1= c_1b_0$ is a constant envelope for $\mathcal{G}_h$
and $\mathcal{G}_h$ is a VC-type class. In addition,
%
\begin{equation}
\sup_{g \in\mathcal{G}_h} \mathbb{E}\bigl[g^2(X_i)
\bigr]\leq h^{d+2} b_1^2\leq
b_1^2
\end{equation}
as $h<1$.
So we can choose $\sigma^2 = h^{d+2} b_1^2$ in Theorem~\ref{thmm::GPA1}.
Applying Theorem~\ref{thmm::GPA1} and \eqref{eq::GP1-2}, there exist
random variables
%
\begin{eqnarray}
\mathbf{B}_1 &\stackrel{d} {=}&\|
\mathbb{B}\|_{\mathcal{G}_h},
\nonumber
\\[-8pt]
\\[-8pt]
\nonumber
\mathbf{B}_2 &\stackrel{d} {=}&\|\mathbb{B}\|_{\mathcal{F}_h}
\end{eqnarray}
such that
%
\begin{eqnarray}\label{eq::GP1-3}
\P \biggl(\bigl |\|\mathbb{G}_n
\|_{\mathcal{G}_h} - \mathbf{B}_1 \bigr|>A_1
\frac{b_1h^{{(d+2)}/{3}}\log^{2/3} n}{\gamma^{1/3}
n^{1/6}} \biggr)&\leq& A_2 \gamma,
\nonumber
\\[-8pt]
\\[-8pt]
\nonumber
\P \biggl( \bigl|\|\mathbb{G}_n\|_{\mathcal{F}_h} - \mathbf{B}_2
\bigr|>A_1\frac{b_1\log^{2/3} n}{\gamma^{1/3}
(nh^{d+2})^{1/6}} \biggr)&\leq& A_2 \gamma
\end{eqnarray}
for two universal constants, when $n$ is sufficiently large and $\gamma
\in(0,1)$.
The second result comes from the one-to-one correspondence between
$\mathcal{G}_h$ and $\mathcal{F}_h$ with a constant scaling.

Now recall \eqref{eq::AH1} from the end of step 1:
\[
\bigl\llvert \sqrt{nh^{d+2}} \Haus(\hat{R}_h,R_h)-
\|\mathbb {G}_n\| _{\mathcal{F}_h}\bigr\rrvert \leq O\bigl(\|
\hat{p}_h-p_h\|^*_{\infty,3}\bigr),
\]
which implies that there exists a constant $C_0>0$ such that for any $a_0>0$,
%
\begin{eqnarray}\label{eq::GP1-Tala}
&& \P \bigl(\bigl\llvert \sqrt{nh^{d+2}} \Haus(
\hat{R}_h,R_h)- \|\mathbb {G}_n
\|_{\mathcal{F}_h}\bigr\rrvert >a_0 \bigr)\nonumber\\
&&\qquad\leq \P \bigl(\|
\hat{p}_h-p_h\|^*_{\infty,3}> C_0\cdot
a_0 \bigr)
\\
&&\qquad\leq4e^{-nh^{d+6}C_1a_0} \nonumber
\end{eqnarray}
for some constant $C_1$ as $n$ is sufficiently large.
Note that we apply Talagrand's inequality (see Lemma~\ref{lem::PB}) in
the last inequality.

Choose $a_0 = \frac{1}{\sqrt{nh^{d+6}}}$ in \eqref{eq::GP1-Tala},
combine it with \eqref{eq::GP1-3} and use the fact that
$\frac{1}{\sqrt{nh^{d+6}}}$ converges much faster than
$\frac{1}{(nh^{d+2})^{1/6}}$, to conclude that
%
\begin{equation}
\P \biggl(\bigl\llvert \sqrt{nh^{d+2}} \Haus(\hat{R}_h,R_h)-
\mathbf{B}_2 \bigr\rrvert >A_3\frac{\log^{2/3} n}{\gamma^{1/3} (nh^{d+2})^{1/6}} \biggr)
\leq A_4 \gamma \label{eq::GP1-4}
\end{equation}
for some constants $A_2,A_4$.
We can replace $A_1$ by $A_3$ and $A_2$ by $A_4$ to absorb the extra
small terms from \eqref{eq::GP1-Tala} and the envelope $b_1$.

\textit{Step \textup{3}---Anti-concentration bound.}

To convert the above result into a Berry--Esseen type bound, we use
the anti-concentration inequality in (Corollary~2.1)
in~\citet{CHERNOZHUKOV2013a}; a similar result appears in
Chernozhukov, Chetverikov and\break =
Kato
(\citeyear{CHERNOZHUKOV2013c,CHERNOZHUKOV2013b}). Here we use a
modification of the anti-concentration inequality.

\begin{lem} [{[Modification of Corollary~2.1 in~\citet{CHERNOZHUKOV2013a}]}]
\label{lem::GPA3}
Let $X_t$ be a Gaussian process with index $t\in T$,
and with semi-metric $d_T$
such that $\mathbb{E}(X_t)= 0, \mathbb{E}(X_t^2)=1$ for all $t\in T$.
Assume that $\sup_{t\in T}X_t<\infty$ a.s. and there exists a
random variable $Y$ such that
$\mathbb{P}(\llvert Y-\sup_{t\in T}|X_t|\rrvert >\eta)<\delta(\eta
)$. If $A(|X|)= \mathbb{E} (\sup_{t\in T} |X_t| )<\infty
$, then
\[
\sup_{t} \Bigl\llvert \P(Y<t) - \P \Bigl(\sup
_{t\in T}|X_t|<t \Bigr) \Bigr\rrvert \leq
A_5 \bigl(\eta+\delta(\eta) \bigr) A\bigl(|X|\bigr)
\]
for some constant $A_5$.
\end{lem}

This lemma is a direct application of Corollary~2.1 of
Chernozhukov, Chetverikov and
Kato (\citeyear
{CHERNOZHUKOV2013a}), so we omit the proof.
We apply Lemma~\ref{lem::GPA3} to equation~\eqref{eq::GP1-4} which yields
%
\begin{eqnarray}
&&\sup_{t} \bigl|\P \bigl(\sqrt{nh^{d+2}}
\Haus(R_h,\hat{R}_h)<t \bigr)- \P \bigl( \|\mathbb{B}
\|_{\mathcal{F}_h}<t \bigr) \bigr|
\nonumber
\\[-8pt]
\\[-8pt]
\nonumber
&&\qquad=A_6 \biggl(A_3\frac{\log^{2/3} n}{\gamma^{1/3} (nh^{d+2})^{1/6}} +A_4
\gamma \biggr),
\end{eqnarray}
where $A_6 = A_5\times A(\|\mathbb{B}\|_{\mathcal{F}_h})<\infty$
is a constant.
We use the fact that $\mathbf{B}_2$ and $\|\mathbb{B}\|_{\mathcal
{F}_h}$ have the same distribution.
Choosing $\gamma= O ((\frac{\log^4 n}{nh^{d+2}})^{1/8}
)$ completes the proof.

For the case of $\dist_{\Pi}(\hat{R}_h,R_h)$, the result follows by
using \eqref{eq::aHaus2}
rather than \eqref{eq::AH1} in the empirical approximation.
\end{pf*}

\begin{pf*}{Proof of Theorem~\ref{thmm::GPB}}
The proof for the bootstrap result is very similar to the previous
theorem. The major difference is that the estimated ridges and the
smoothed ridges have different supports. This makes the functional
spaces different. Our strategy for proving this theorem
has three steps. First, we show that the Hausdorff distance
$\Haus(\hat{R}^*_h, \hat{R}_h)$ conditioned on the observed data can be
approximated by an empirical process. This is the same as step 1 in
proof of Theorem~\ref{thmm::GP}. Second, we apply the result of
Theorem~\ref{thmm::GP} to bound the difference between the distributions
of $\Haus(\hat{R}^*_h, \hat{R}_h)$ and a Gaussian process defined on the
$\hat{R}_h$. This uses the second and the third steps of the
previous proof. The last step shows that the Gaussian process
defined on $\hat{R}_h$ is asymptotically the same as being defined
on~$R_h$.
Let
$\mathbb{X}_n\in\mathcal{X}_n$,
and recall that
by Lemma~\ref{lem::PB},
$\P(\mathcal{X}_n)>1-5e^{-nh^{d+8}D_1}$.

\textit{Step \textup{1}---Empirical approximation.}

Let $\mathbb{X}_n = \{X_1,\ldots,X_n\}$ be the observed data.
Let $\mathbb{G}_n^*(\mathbb{X}_n) = \sqrt{n}(\P^*_n-\P_n)$.
Let $\hat{p}^*_h$ be the bootstrap KDE (KDE
based on the bootstrap sample).

In the following, we assume that $\mathbb{X}_n\in\mathcal{X}_n$ and
treat $\mathbb{X}_n$ as fixed. Hence, $\hat{p}_h$ and $\hat{R}_h$ are
fixed. In this case, Theorem~\ref{LU2} can be applied to the local
uncertainty vector, that is,
%
\begin{equation}
\bigl\|\mathbf{d}\bigl(x, \hat{R}^*_h\bigr) -\mathbb{G}^*_n(
\mathbb {X}_n) (\hat {f}_{n,x})\bigr\| = \bigl\|\hat{p}^*_h
-\hat{p}_h\bigr\|^*_{\infty
,3},\qquad x\in\hat{R}_h,
\end{equation}
where
%
\begin{equation}
y\mapsto\hat{f}_{n,x}(y) = \frac{1}{\sqrt{h^{d+2}}} \hat{N}_n(x)
\hat{H}_{N,n}(x)^{-1} \hat{N}_n(x)^T
(\nabla K) \biggl(\frac
{x-y}{h} \biggr)\in\mathbb{R}^{d}.
\label{eq::GPB-1}
\end{equation}
Note that $\hat{N}_n(x)$ is the matrix with column space equal to the
normal space
of $\hat{R}_h$ at $x$, and $\hat{H}_{N,n}(x)$ is the
corresponding subspace Hessian matrix of the space spanned by columns
of $\hat{N}_n(x)$.
Define
%
\begin{equation}
\hat{\mathcal{F}_h}(\mathbb{X}_n) = \bigl\{
\mathbf{w}^T \hat {f}_{n,x}\dvtx\mathbf{w}\in
\mathbb{R}^{d}, \|\mathbf{w}\|=1, x \in \hat {R}_h\bigr\}.
\label{eq::GPB-2}
\end{equation}
Then by the same argument as in the paragraph before the proof of
Theorem~\ref{thmm::GP}, we have a similar result to \eqref{eq::AH1},
%
\begin{equation}
\bigl\llvert \Haus\bigl(\hat{R}^*_h, \hat{R}_h\bigr)-
\bigl\|\mathbb{G}^*_n(\mathbb {X}_n)\bigr\|_{\hat{\mathcal{F}_h}(\mathbb{X}_n)}\bigr
\rrvert = O \bigl(\bigl\| \hat{p}^*_h -\hat{p}_h
\bigr\|^*_{\infty,3} \bigr). \label{eq::GPB-3}
\end{equation}

\textit{Step \textup{2}---Gaussian approximation.}

We use the same proof
as in Theorem~\ref{thmm::GP}.
We apply Theorem~\ref{thmm::GP} to conclude that
%
\begin{eqnarray}\label{eq::GPB-4}
&& \sup_{t} \bigl|\P \bigl(\sqrt{nh^{d+2}}
\Haus\bigl(\hat{R}^*_h,\hat{R}_h\bigr)<t |
\mathbb{X}_n \bigr)- \P\bigl ( \|\mathbb{B}\|_{\hat{\mathcal{F}}_h(\mathbb
{X}_n)}<t|
\mathbb{X}_n \bigr) \bigr|
\nonumber
\\[-8pt]
\\[-8pt]
\nonumber
&&\qquad=O \biggl( \biggl(\frac{\log^4 n}{nh^{d+2}} \biggr)^{1/8} \biggr).
\end{eqnarray}

\textit{Step \textup{3}---Support approximation.}

In the previous step, the approximating distribution is a Gaussian
process over the function space $\hat{\mathcal{F}}_h(\mathbb{X}_n)$,
which is not the same as $\mathcal{F}_h$. Now we apply
Lemma~\ref{lem::GCR} and the fact that $\|\hat{p}_h-p\|^*_{\infty
,2}/h=O(\|\hat{p}_h-p\|^*_{\infty,3})$ to get
%
\begin{equation}
 \sup_{t} \bigl\llvert \P \bigl( \|\mathbb{B}
\|_{\hat{\mathcal{F}}_h(\mathbb
{X}_n)}<t |\mathbb{X}_n \bigr)- \P \bigl( \|\mathbb{B}
\|_{\mathcal{F}_h}<t \bigr)\bigr\rrvert =O \bigl(\bigl(\|\hat{p}_h-p
\|^*_{\infty,3}\bigr)^{1/3} \bigr).
\label{eq::GPB-5}
\end{equation}
Combining \eqref{eq::GPB-4}, \eqref{eq::GPB-5} and the fact that
$\|\hat{p}_h-p\|^*_{\infty,3}=O(\frac{\log n}{nh^{d+6}})$, we conclude
\begin{eqnarray*}
&&\sup_{t} \bigl\llvert \P \bigl(\sqrt{nh^{d+2}}
\Haus\bigl(\hat{R}^*_h,\hat {R}_h\bigr)<t |
\mathbb{X}_n \bigr) - \P \bigl(\|\mathbb{B}\|_{\mathcal{F}_h} < t \bigr)\bigr
\rrvert
\\
&&\qquad = O \biggl(\frac{\log n}{ (nh^{d+2} )^{1/8}} \biggr)+ O \biggl( \biggl(\frac{\log n}{nh^{d+6}}
\biggr)^{1/6} \biggr).
\end{eqnarray*}
\upqed\end{pf*}

Consider two densities $p_1,p_2$ satisfying conditions (A1), (P1)--(P2).
Let $R_1,R_2$ be the density ridges for $p_1,p_2$, respectively.
We assume conditions (K1)--(K2).
Define
%
\begin{equation}
\mathcal{F}_k =\bigl\{\mathbf{w}^T f_{x,k}
\dvtx\mathbf{w}\in\mathbb{R}^d, \|\mathbf{w}\|=1, x\in R_k
\bigr\},\qquad k=1,2, \label{eq::GCR1}
\end{equation}
where
%
\begin{equation}
f_{x,k} = \frac{1}{\sqrt{h^{d+2}}}N_k(x) H_{N,k}(x)^{-1}
N_k(x)^T (\nabla K) \biggl(\frac{x-y}{h} \biggr) \in
\mathbb{R}^d \label{eq::GCR2}.
\end{equation}
Note that we have two indices $x,\mathbf{w}$ for each element in
$\mathcal{F}_1$ and $\mathcal{F}_2$. The first index $x$ is the
\emph{location}, and the second index $\mathbf{w}$ is the
\emph{direction}.
$N_k(x)$ is the normal matrix (as $x\in
R_k$, its column space is the normal space) defined by
Lemma~\ref{lem::NS} at $x$, and $H_{N,k}(x)$ is the subspace Hessian in
the columns space of $N_k(x)$.

\begin{lem}[(Gaussian comparison on two ridges)]
\label{lem::GCR}
When
$\|p_1-p_2\|^*_{\infty,3}$ is sufficiently small, we have
\[
\sup_{t} \bigl\llvert \P\bigl(\|\mathbb{B}\|_{\mathcal{F}_1} < t
\bigr) - \P \bigl(\|\mathbb{B}\|_{\mathcal{F}_2} < t \bigr) \bigr\rrvert = O \bigl(\bigl(
\|p_1-p_2\|^*_{\infty,3}+\Haus(R_1,R_2)/h
\bigr)^{1/3} \bigr).
\]
\end{lem}

The proof can be found in the supplementary material [\citet{chen2014ridgesupp}].

\section*{Acknowledgments}
We thank the reviewers for helpful comments.
We also thank Alessandro Rinaldo for useful comments
about Lemma~\ref{LUlem3}.

\begin{supplement}[id=suppA]
\stitle{Supplementary proofs: Asymptotic theory for density ridges\\}
\slink[doi]{10.1214/15-AOS1329SUPP} 
\sdatatype{.pdf}
\sfilename{aos1329\_supp.pdf}
\sdescription{The supplementary material contains proofs of
Lemmas~\ref{lem::NS0},~\ref{lem::NS},~\ref{lem::SH},~\ref
{LUlem3},~\ref{lem::PB},
\ref{lem::AH},~\ref{lem::GCR}.}
\end{supplement}

%

%
%

%



\printaddresses
\end{document}